\documentclass[twocolumn,aps,prd,showpacs,nofootinbib,superscriptaddress,
preprintnumbers,floatfix,10pt]{revtex4-1}
\usepackage{amsmath}
\usepackage{amsfonts}
\usepackage{amssymb}
\usepackage{epsfig}
\usepackage{graphicx}
\usepackage{color}
\usepackage{slashed}
\usepackage{epstopdf}
\usepackage{dsfont}
\usepackage{mathrsfs}

\usepackage{hyperref}

\newcommand{\Eqref}[1]{Eq.~\eqref{#1}}

\begin{document}

\preprint{}

\title{Renormalization Flow of Nonlinear Electrodynamics} 

\author{Holger Gies}
\email{holger.gies@uni-jena.de}
\affiliation{Theoretisch-Physikalisches Institut, 
Abbe Center of Photonics, Friedrich Schiller University Jena, Max Wien 
Platz 1, 07743 Jena, Germany}
\affiliation{Helmholtz-Institut Jena, Fr\"obelstieg 3, D-07743 Jena, Germany}
\affiliation{GSI Helmholtzzentrum für Schwerionenforschung, Planckstr. 1, 
64291 Darmstadt, Germany} 
\author{Julian Schirrmeister}
\email{julian.schirrmeister@uni-jena.de}
\affiliation{Theoretisch-Physikalisches Institut, 
Abbe Center of Photonics, Friedrich Schiller University Jena, Max Wien 
Platz 1, 07743 Jena, Germany}

\begin{abstract}
We study the renormalization flow of generic actions that depend on the 
invariants of the field strength tensor of an abelian gauge field. While the 
Maxwell action defines a Gaussian fixed point, we search for 
further non-Gaussian fixed points or rather fixed functions, i.e., globally 
existing Lagrangians of the invariants. Using standard small-field expansion techniques for the resulting functional flow equation, a large number of fixed points is obtained, which -- in analogy to recent findings for a shift-symmetric scalar field -- we consider as approximation artifacts.

For the construction of a globally existing fixed function, we pay attention to the use of proper initial conditions. Parametrizing the latter by the photon anomalous dimension, both the coefficients of the weak-field expansion are fully determined and those of the large-field expansion can be matched such that a global fixed function can be constructed for magnetic fields. The anomalous dimension also governs the strong-field limit. Our results provide evidence for the existence of a continuum of non-Gaussian fixed points parametrized by a small positive anomalous dimension below a critical value.

We discuss the implications of this result within various scenarios with and without additional matter. For the strong-field limit of the 1PI 
QED effective action, where the anomalous dimension is determined by electronic fluctuations, our result suggests the existence of a singularity free strong-field limit, circumventing the standard conclusions connected to 
the perturbative Landau pole.  

\end{abstract}

\maketitle
\section{Introduction}
\label{intro}
Relativistic models of nonlinear electrodynamics have an extensive history in 
field theory, beginning with Born- or Born-Infeld theory motivated by the 
removal of the divergence of the electron's self-energy in a classical setting 
\cite{Born:1933qff, Born:1933pep,Born:1934gh}, reemerging also as an effective 
theory of the open string \cite{Fradkin:1985qd}. The Heisenberg-Euler 
theory \cite{Heisenberg:1936nmg, Weisskopf:1936hya, Schwinger:1951nm, 
Dunne:2004nc} represents not only the presumably correct theory of the nonlinear 
response of the electrodynamic quantum vacuum according to quantum 
electrodynamics (QED), but is also a hallmark of the concept of effective field 
theory now being ubiquitous in quantum field theory. Discovering the plethora of 
phenomena predicted by the Heisenberg-Euler action \cite{Dittrich:2000zu,Marklund:2008gj,DiPiazza:2011tq,Fedotov:2022ely} is currently a substantial 
research endeavor in strong-field physics.   

Perturbative renormalizability arguments suggest that nonlinear models of 
electrodynamics should not be viewed as a fundamental 
theory, as the nonlinear interactions are power-counting nonrenormalizable. 
Whether or not the naive perturbative conclusion can be extended to a strong 
coupling region is explored in the present work. In principle, perturbative 
arguments fail in presence of non-Gaussian fixed points which are a prerequisite 
for the construction of high-energy complete theories based on the concept of 
asymptotic safety \cite{Weinberg:1976xy,Weinberg:1980gg}.  

Ultraviolet (UV) completeness of QED, i.e., including the fluctuations of
fermionic or scalar charged particles, has been at the center of interest since
the discovery of the perturbative Landau pole
\cite{Landau:1955,GellMann:1954fq,Johnson:1964da}. Though the Landau pole
divergence of the coupling may or may not be an artifact of perturbation theory,
there is consensus among various methods that the strong-coupling regime of
conventional QED cannot be connected to the weak-coupling regime realized in
nature because of chiral symmetry breaking and mass generation
\cite{Miransky:1984ef,Gockeler:1997dn,Kim:2000rr,Aoki:1996fh,Gies:2004hy}.
Beyond the conventional scenarios, UV completion in the large-flavor number
limit \cite{PalanquesMestre:1983zy,Gracey:1996he,Shrock:2013cca,Antipin:2017ebo,
Antipin:2018zdg,Dondi:2019ivp,Dondi:2020qfj}, at a non-Gaussian Pauli coupling
fixed point \cite{Djukanovic:2017thn,Gies:2020xuh,Gies:2022aiz,Gies:2024ugz}, novel 
resummation schemes \cite{Evans:2023mxp}, or UV completion mediated by
gravitational fluctuations \cite{Harst:2011zx,Christiansen:2017gtg} have
recently been discussed.

At first sight, it is thus not surprising that perturbative renormalization
group (RG) resummations also find a Landau pole divergence in the strong-field
limit of the higher-loop resummed Heisenberg-Euler effective action
\cite{Ritus:1977iu,Ritus:1977iu,Dittrich:1985yb}. This holds both for the
effective action as the 1PI generating functional, as well as for the Schwinger
functional that include 1PR resummations
\cite{Gies:2016yaa,Karbstein:2019wmj,Karbstein:2021gdi,Evans:2023jqy}. At second
glance, such a strong-field divergence may appear less plausible at least for
the case of a homogeneous magnetic background. This is because the latter does
not transfer energy to charged fluctuations and thus should not \textit{per se}
probe the high-momentum regime where Landau-pole singularities may play a role. 

In the present work, we address the question of a possible existence of
UV-complete nonlinear electrodynamics (without further charged matter degrees of
freedom) as well as the strong-field limit of Heisenberg-Euler-type theories
(with a minimum charged matter content), using methods of functional
renormalization. More specifically, we derive the general nonperturbative RG
flow equation for action functionals depending on the gauge- and
Lorentz-invariant combinations of the field strength. For both aspects, we find
that the criterion of \textit{global existence} of functions satisfying the
fixed-point equation for the action is most relevant. This is, in fact, familiar
from technically similar searches for Wilson-Fisher-type fixed points
\cite{Hasenfratz:1985dm,Morris:1998da,Bervillier:2007tc, Codello:2012ec,
Hellwig:2015woa, Borchardt:2015rxa, Dietz:2015owa,
Litim:2016hlb,Juttner:2017cpr}, scaling solutions in fermionic/Yukawa theories
\cite{Vacca:2015nta,Knorr:2016sfs,Dabelow:2019sty}, UV completions of gauged
Higgs models \cite{Gies:2015lia,Gies:2016kkk} or gauged Yukawa models
\cite{Gies:2018vwk,Gies:2019nij,Schreyer:2020}, and studies in quantum gravity
\cite{Demmel:2015oqa, Borchardt:2015rxa,Ohta:2015efa,Morris:2022btf}. 

As for the quest for matterless UV-complete asymptotically safe nonlinear
electrodynamics, our answer is in the negative as long as standard search
methods for fixed points based on \textit{improper} initial conditions are used.
For this case, our results are rather similar to analogous studies of
shift-symmetric scalar theories or nonlinear abelian models based on the Maxwell
invariant \cite{Laporte:2022ziz,deBrito:2023myf}. While the weak-field expansion
finds many potential fixed-point candidates similar to \cite{Laporte:2022ziz},
the picture is quite comparable to the shift-symmetric scalar model, where the eigenperturbations around the fixed points are not integrable with respect to the induced measure \cite{deBrito:2023myf}; therefore, only the trivial Gaussian fixed point remains in this analysis.

By contrast, we do find globally existing fixed-point actions, once the
construction is based on \textit{proper} initial conditions. We parametrize the latter in
terms of the anomalous dimension of the photonic field which is either a free
parameter, or could effectively be provided by charged matter fluctuations. With
this parameter, we are able to construct a global action in the direction of one
of the invariants by a nontrivial matching of the small- and large-field
expansions. The approximations involved can be applied to the case of a
purely magnetic field and thus provide evidence for the absence of
Landau-pole-type singularities in the strong-field limit of this type. 

The paper is structured as follows: in Sect. \ref{sec:two}, we introduce
the setting for general theories of nonlinear electrodynamics including
Minkowskian as well as Euclidean formulations. In Sect. \ref{sec:three}, we
derive the RG flow on the considered theory space using the functional
renormalization group. On a fixed point, the resulting flow equation reduces to
a fixed-function equation, defining scaling solutions for generic effective
Lagrangians. We also motivate and substantiate a set of approximations which
simplify the analysis of the differential equation. Section \ref{sec:four} is
devoted to a standard analysis of the (reduced) fixed-function equation
including the critical regime based on a conventional small-field expansion.
Whereas this analysis corresponds to improper initial conditions,
Sect.~\ref{sec:five} investigates the fixed-function equation using proper
initial conditions. A one-parameter family of global solutions are constructed
on small- and large-field expansions for small positive anomalous dimensions.
Our approximations are checked in Sect.~\ref{sec:six} by tackling the full
partial differential equation in the small field regime. We interpret our
results in the light of various scenarios in Sect.~\ref{sec:seven} and conclude
in Sect.~\ref{sec:conc}.

\section{Nonlinear Electrodynamics}
\label{sec:two}
Maxwell's theory of electrodynamics (ED) in vacuum is a linear theory entailing
a strict superposition principle. It can be defined in terms of the gauge
potential $(\bar{\mathsf{A}}_{\mu})$ in four-dimensional Minkowski space and the
corresponding field strength tensor $(\bar{\mathsf{F}}_{\mu\nu})$ with its
components being connected to the gauge potential in the usual way,
$\bar{\mathsf{F}}_{\mu\nu} = \partial_{\mu}\bar{\mathsf{A}}_{\nu} -
\partial_{\nu}\bar{\mathsf{A}}_{\mu}$.\footnote{For the gauge field, the field
strength, and the action, we consistently use a notation, where serifless fonts
are used for Minkowski-valued quantities. The standard notation is reserved for
the renormalized quantities on Euclidean space, which will be defined later. An
overbar indicates an unrenormalized and typically dimensionful quantity.} Using
the gauge and Lorentz invariant scalars formed from the field strength tensor
and its Hodge dual $\left((\star\bar{\mathsf{F}})^{\mu\nu}\right)$, that is
\begin{equation}
 \bar{\mathcal{F}} := 
\frac{1}{4}\bar{\mathsf{F}}_{\mu\nu}\bar{\mathsf{F}}^{\mu\nu}, \hspace*{0.15cm} 
\bar{\mathcal{G}} := 
\frac{1}{4}\bar{\mathsf{F}}_{\mu\nu}(\star \bar{\mathsf{F}})^{\mu\nu} =
\frac{1}{8} \varepsilon^{\mu\nu\rho\sigma} \bar{\mathsf{F}}_{\mu\nu} \bar{\mathsf{F}}_{\rho\sigma}
\label{eq:b.c}
\end{equation}
(using the convention $\varepsilon^{0123}=1$), the free action reads
\begin{equation}
\mathsf{S}[\bar{\mathsf{A}}] = 
\int\limits_{\mathbb{R}^{3,1}}-\bar{\mathcal{F}}\bigl(\bar{\mathsf{A}}(x)\bigr)\,\mathrm{d}^{4}x.
\label{eq:b.b}
\end{equation}
Further local invariants involve derivatives of the field strength. 

The most general effective action functional $\Gamma$ of nonlinear
electrodynamics may depend on all possible invariants; a generic theory can 
thus be parametrized by a local Lagrangian depending on the field strength and 
its derivatives:
\begin{equation}
\mathsf{\Gamma}[\bar{\mathsf{A}}] = 
\int\limits_{\mathbb{R}^{3,1}}\bar{\mathcal{L}}\bigl(\bar{\mathsf{F}}(x), 
\partial_{\mu} \bar{\mathsf{F}}(x),\partial_{\nu}\partial_{\lambda} 
\bar{\mathsf{F}}(x),\ldots \bigr) \, \mathrm{d}^{4}x.
\label{eq:b.d}
\end{equation}
In the present work, we ignore possible dependencies on the derivative terms 
and concentrate on the full functional dependence on the two invariants 
$\bar{\mathcal{F}}$ and $\bar{\mathcal{G}}$. This may be viewed as the leading order 
of a systematic derivative expansion of the action 
\cite{Mamaev:1981dt,Gusynin:1998bt,Cangemi:1994by,Fliegner:1997rk,Dittrich:1996dz,Karbstein:2021obd} in the spirit of the Heisenberg-Euler 
expansion \cite{Heisenberg:1936nmg}. However, in contradistinction to 
conventional derivative expansions where higher-order derivatives have to be 
small compared to a physical mass scale, our expansion is based on a comparison 
to a running RG scale $k$. The validity criterion therefore is that the 
derivative terms should have a small influence on the flow 
of the nonderivative terms at any scale $k$; they do not necessarily have to 
be numerically small. We thus approximate the general Lagrangian by an
$\bar{\mathsf{F}}$-dependent function, or equivalently a function 
of the invariants, $\bar{\mathcal{L}}(\bar{\mathsf{F}}, 
\partial_{\mu} \bar{\mathsf{F}},\ldots ) \approx \bar{\mathcal{L}}
(\bar{\mathsf{F}},0,\ldots)
\equiv \bar{\mathcal{W}}\left(\bar{\mathcal{F}}(\bar{\mathsf{F}}),
\bar{\mathcal{G}}(\bar{\mathsf{F}})\right)$, reducing the 
action to 
\begin{equation}
\mathsf{\Gamma}[\bar{\mathsf{A}}] \approx
\int\limits_{\mathbb{R}^{3,1}}\bar{\mathcal{W}}\Bigl(\bar{\mathcal{F}}(\bar{\mathsf{A}}(x)),
\bar{\mathcal{
G}}(\bar{\mathsf{A}}(x))\Bigr)\, \mathrm{d}^{4}x.
\label{eq:b.f}
\end{equation}
This defines the class of action functionals covering nonlinear 
generalizations of vacuum electrodynamics to leading-derivative order. The corresponding equations of 
motion generically represent hyperbolic second-order nonlinear partial 
differential equations for which initial-value problems can be formulated.

In the following, we restrict ourselves to parity-invariant theories. Since 
$\bar{\mathcal{G}}$ is parity-odd, $\bar{\mathcal{W}}$ should be considered as 
an even function of $\bar{\mathcal{G}}$, i.e., instead of
$\bar{\mathcal{W}}(\bar{\mathcal{F}},\bar{\mathcal{G}})$ we write $\bar{\mathcal{W}}(\bar{\mathcal{F}},\bar{\mathcal{G}}^2)$.

As our renormalization group analysis will be performed in Euclidean spacetime, 
let us detail the connection between Minkowski-valued and Euclidean quantities: 
In $d=4$ dimensional Minkowski space, the components of the
antisymmetric field strength tensor, $\bar{\mathsf{F}}_{\mu\nu}$, are related to the 
electric and magnetic field components by $\bar{\mathsf{F}}_{0i} = 
\bar{\mathsf{E}}_{i}$ and $\bar{\mathsf{F}}_{ij} = 
\varepsilon_{ijl}\bar{\mathsf{B}}_{l}$. In terms of the fields, the invariant 
scalars read $\bar{\mathcal{F}}= 
\frac{1}{2}(\bar{\mathsf{B}}^{2}-\bar{\mathsf{E}}^{2})$ and 
$\bar{\mathcal{G}}= 
- \bar{\mathsf{E}} \cdot \bar{\mathsf{B}}$.

In the Euclidean, we start from a Euclidean gauge potential $(\bar{A}_{\mu})$ with field strength components $\bar{F}_{\mu\nu} = \partial_{\mu}\bar{A}_{\nu} - \partial_{\nu}\bar{A}_{\mu}$ and the components of its Hodge dual $(\star\bar{F})^{\mu\nu} = \frac{1}{2}\varepsilon^{\mu\nu\rho\sigma}\bar{F}_{\rho\sigma}$. The corresponding invariants read
$\bar{\mathscr{F}} := \frac{1}{4}\bar{F}_{\mu\nu}\bar{F}^{\mu\nu}$ and $\bar{\mathscr{G}} := \frac{1}{4}\bar{F}_{\mu\nu}(\star\bar{F})^{\mu\nu}$, where the
Euclidean metric is used for the contractions. Identifying the Euclidean field strength components as $\bar{F}_{0i} = \bar{E}_{i}$ and $\bar{F}_{ij} = \varepsilon_{ijl}\bar{B}_{l}$, we obtain for the invariants $\bar{\mathscr{F}} = \frac{1}{2}(\bar{B}^{2} + 
\bar{E}^{2})$ and $\bar{\mathscr{G}} = \bar{E}\cdot\bar{B}$.

On the level of the invariants $\bar{\mathcal{F}}$ and
$\bar{\mathscr{F}}$, the transition between Minkowskian and Euclidean
spacetime is captured by the field replacement rule
$\left(\bar{\mathsf{E}},\bar{\mathsf{B}}\right) \leftrightarrow \left(-\imath\bar{E},\bar{B}\right)$.
This also implies a relation between $\bar{\mathcal{G}}$ and $\bar{\mathscr{G}}$,
which in total yields: $\bar{\mathcal{F}} \leftrightarrow \bar{\mathscr{F}}$ and
$\bar{\mathcal{G}} \leftrightarrow \imath\bar{\mathscr{G}}$.

Including the Wick rotation in coordinate space, the corresponding Euclidean 
action, e.g., of Maxwell's theory reads
\begin{equation}
S[\bar{A}] = 
\int\limits_{\mathbb{R}^{4}}\bar{\mathscr{F}}\bigl(\bar{A}(x)\bigr)\,\mathrm{d}^{4}x.
\label{eq:b.g}
\end{equation}
Analogously to \Eqref{eq:b.f}, the corresponding class of general nonlinear 
theories of electrodynamics investigated in this work is described by a 
Euclidean action
\begin{equation}
\Gamma[\bar{A}] := 
\int\limits_{\mathbb{R}^{4}}\bar{\mathscr{W}}\Bigl(\bar{\mathscr{F}}(\bar{A}(x)),\bar{
\mathscr{G}}(\bar{A}(x))^{2}\Bigr)\,\mathrm{d}^{4}x,
\label{eq:b.h}
\end{equation}
where the function $\bar{\mathscr{W}}$ is the Euclidean analog of $\bar{\mathcal{W}}$.
On the level of the Lagrangian, the transition from Euclidean back to
Minkowskian spacetime is thus performed by the replacements 
$\bar{\mathscr{W}}\to- \bar{\mathcal{W}}$, $\bar{\mathscr{F}}\to 
\bar{\mathcal{F}}$, and $\bar{\mathscr{G}}\to -\imath\bar{\mathcal{G}}$.
The latter implies $\bar{\mathscr{G}}^{2} \to -\bar{\mathcal{G}}^{2}$.

\section{RG Flow and Fixed Functions}
\label{sec:three}

Let us list the ingredients for our functional RG analysis for theories
based on the action in \Eqref{eq:b.h}. For conceptual and technical details, we refer the reader to reviews on the functional RG \cite{Berges:2000ew,Pawlowski:2005xe,Gies:2006wv,Delamotte:2007pf,Rosten:2010vm,Braun:2011pp,Dupuis:2020fhh}.

\subsection{Scale-Dependent Effective Action}
\label{par:threeone}

Using the functional RG, we quantize nonlinear electrodynamics in a Wilsonian sense momentum shell by momentum shell. Quantization over a finite amount of scales is always possible in the spirit of an effective field theory. In addition, we intend to search for fixed points, or rather fixed functions, of the RG flow that have the potential to allow for a consistent 
quantization on all scales. For this, we use the Wetterich 
equation~\cite{Wetterich:1992yh,Ellwanger:1993mw,Morris:1993qb,Bonini:1992vh} for a scale-dependent one-particle irreducible (1PI) effective action $\Gamma_k$,
\begin{equation}
k\partial_{k}\Gamma_{k}[\bar{A}] = 
\frac{1}{2}\mathbf{Tr}\left[\left(\Gamma_{k}^{(2)} + 
\mathcal{R}_{k}\right)^{-1}k\partial_{k}\mathcal{R}_{k}\right][\bar{A}],
\label{eq:b.j}
\end{equation}
where $\Gamma_{k}^{(2)}$ denotes the second functional derivative of 
$\Gamma_{k}$ with respect to the gauge field $\bar{A}$. The quantity 
$\mathcal{R}_{k}$ is a regulator that controls infrared (IR) mode suppression below a 
momentum scale $k$ and implements the Wilsonian momentum-shell integration. For 
a given initial condition, e.g., at a UV scale $\Gamma_{k=\Lambda}$, the action 
$\Gamma_{k=0}$ includes all quantum fluctuations with momenta below $\Lambda$ 
\cite{Berges:2000ew,Pawlowski:2005xe,Gies:2006wv,Delamotte:2007pf,Rosten:2010vm,Braun:2011pp,Dupuis:2020fhh}.

In the present work, we parametrize the action functional $\Gamma_k$ by a 
scale-dependent variant of the nonlinear theory space spanned by
\Eqref{eq:b.h} amended by a Lorenz gauge-fixing term,
\begin{equation}
\Gamma_{k}[\bar{A}] := 
\int\limits_{\mathbb{R}^{4}}\left(\bar{\mathscr{W}}_{k}\left(\bar{\mathscr{F}},
\bar{\mathscr{G}}^{2}\right) + 
\frac{1}{2\alpha}Z_{k}\left(\partial_{\mu}\bar{A}^{\mu}\right)^{2}\right)\,
\mathrm{d}^{4}x,
\label{eq:c.c}
\end{equation}
where we have suppressed the $x$ dependencies.

We assume that the function $\bar{\mathscr{W}}_k$ features a weak-field 
expansion of the form
$\bar{\mathscr{W}}_{k}\left(\bar{\mathscr{F}},\bar{\mathscr{G}}^{2}\right) 
= Z_k \bar{\mathscr{F}} + \ldots$, where $Z_k$ can be 
interpreted as a wave function renormalization.
We have included $Z_k$ also in 
the gauge-fixing term in order to obtain a standard form of the gauge-fixed 
propagator including the gauge-fixing parameter $\alpha\in\mathbb{R}$. Analogous parametrizations of the effective action also including the nonabelian case have been studied in \cite{Reuter:1993kw,Reuter:1997gx,Gies:2002af,Gies:2003ic,Gies:2004hy}. 

The scale dependence of $Z_k$ is encoded in the anomalous dimension of the 
gauge field,
\begin{equation}
\eta_{k} := -k\partial_{k}\ln\left(Z_{k}\right).
\label{eq:c.ca}
\end{equation}

For the analysis of the RG flow, it is useful to introduce dimensionless 
renormalized quantities. In $d=4$ dimensions, the corresponding rescalings using the scale 
$k$ read:
\begin{equation}
\mathscr{F} := Z_{k}k^{-4}\bar{\mathscr{F}}, \quad \mathscr{G} := 
Z_{k}k^{-4}\bar{\mathscr{G}}, \quad  w_{k} := k^{-4}\bar{\mathscr{W}}_{k}.
\label{eq:c.cb}
\end{equation}
The $Z_k$ rescaling of the field implies that the weak-field expansion of 
the \textit{field-strength potential} $w_{k}$ starts with $w_k = \mathscr{F} + \dots$.
In accordance with \Eqref{eq:c.cb}, we also introduce a 
dimensionless-renormalized field strength and a conveniently rescaled 
(though dimensionful) gauge field:
\begin{equation}
F := \sqrt{Z_{k}}k^{-2}\bar{F}, \quad A := \sqrt{Z_{k}}k^{-2}\bar{A}.
\label{eq:c.cc}
\end{equation}
Note that $A$ carries an inverse mass dimension,  
such that the dimensionless field strength components $F_{\mu\nu}$ maintain their standard form, $F_{\mu\nu} = 
\partial_{\mu}A_{\nu} - \partial_{\nu}A_{\mu}$, and the scale-dependent effective action yields
\begin{equation}
\Gamma_{k}[A] = 
k^{4}\int\limits_{\mathbb{R}^{4}}\left(w_{k}(\mathscr{F},\mathscr{G}^{2}) + 
\frac{1}{2\alpha}(\partial_{\mu}A^{\mu})^{2}\right)\,\mathrm{d}^{4}x.
\label{eq:flowacteucl}
\end{equation}
Here, we have once again suppressed the $x$ dependencies under the integral.

\subsection{RG Flow Equation}
\label{par:threetwo}
For the evaluation of the Wetterich equation, we need the Hessian of the action:
\begin{equation}
\bigl(\Gamma_{k}^{(2)}\bigr)^{\mu\nu}[\bar{A}(A)](x,x^{\prime}) = 
Z_{k}k^{-4}\frac{\delta^{2}\Gamma_{k}[A]}{\delta A_{\mu}(x)\delta A_{\nu}(x^{\prime})}.
\label{eq:c.d}
\end{equation}
With respect to the continuous part of the spectrum, the Hessian can be 
diagonalized in momentum space, since it suffices to consider a homogeneous field strength in order to extract information about the flow of $w_k$.
Then, using \Eqref{eq:flowacteucl}, the following decomposition of $\Gamma_{k}^{(2)}$ in terms of projectors, 
i.e., idempotent endomorphisms, acting on the Lorentz components in field 
space is useful:
\begin{equation}
\Gamma_{k}^{(2)} = Z_{k}k^{2}\left(X_{k}^{T}\mathbf{P}_{T} + X_{k}^{L}\mathbf{P}_{L} + 
\sum_{a=1}^{4}X_{k}^{a}\mathbf{P}_{a}\right).
\label{eq:c.e}
\end{equation}
The projection operators and their coefficients are listed in Table~\ref{tab:c.a}, using the shorthand 
notation
\begin{equation}
w_{k}^{\prime} := \frac{\partial w_{k}}{\partial\mathscr{F}} \qquad \text{and} \qquad
\dot{w}_{k} := \frac{\partial w_{k}}{\partial\left(\mathscr{G}^{2}\right)}.
\label{eq:c.f}
\end{equation}
For convenience, we have also introduced a dimensionless momentum space coordinate $y := p/k$.

Let us elucidate some properties of the field space projection endomorphisms.
The projectors act as linear operators on the space of gauge fields,
that is the space of 1-form fields $\mathfrak{X}^{*}(\mathbb{R}^{4})$
which contains smooth sections of the cotangent bundle
$T^{*}\mathbb{R}^{4}$ with respect
to the standard smooth structure on four-dimensional Euclidean space.
The images of the
projections are linear subspaces of 
$\mathfrak{X}^{*}(\mathbb{R}^{4})$. In particular, $\mathbf{P}_{T}(\mathfrak{X}^{*}(\mathbb{R}^{4}))$ and 
$\mathbf{P}_{L}(\mathfrak{X}^{*}(\mathbb{R}^{4}))$ define a transversal and 
longitudinal component of $\mathfrak{X}^{*}(\mathbb{R}^{4})$, respectively. They are 
orthogonal complements to each other, i.e., $\mathbf{P}_{T} \circ 
\mathbf{P}_{L} = \mathbf{P}_{L}\circ\mathbf{P}_{T} = 0$.
In fact, the corresponding involution $i_{T} := \mathds{1} - 2\mathbf{P}_{T}$ induces a natural
$\mathbb{Z}_{2}$-gradation for $\mathfrak{X}^{*}(\mathbb{R}^{4})$. Since $\mathds{1} = \mathbf{P}_{T} + \mathbf{P}_{L}$,
an arbitrary element $A \in \mathfrak{X}^{*}(\mathbb{R}^{4})$ can thus be decomposed
into transversal and longitudinal parts:
\begin{equation}
A = \mathds{1}(A) = \mathbf{P}_{T}(A) + \mathbf{P}_{L}(A) \equiv A_{T} + A_{L}.
\label{eq:c.g}
\end{equation}
\begin{table}[t]
\centering
\renewcommand{\arraystretch}{2}
\begin{tabular}{|c|c|}
\hline
Projector & Coefficient \\
\hline\hline
$\mathbf{P}_{T} = \mathds{1} - \frac{p\otimes p}{p^{2}}$ & $X_{k}^{T} = 
w_{k}^{\prime}y^{2}$ \\
\hline
$\mathbf{P}_{L} = \frac{p\otimes p}{p^{2}}$ & $X_{k}^{L} = 
\frac{1}{\alpha}y^{2}$ \\
\hline
$\mathbf{P}_{1} = \frac{(Fp)\otimes (Fp)}{(Fp)^{2}}$ & $X_{k}^{1} = 
w_{k}^{\prime\prime}(Fy)^{2}$ \\
\hline
$\mathbf{P}_{2} = \frac{(\star Fp)\otimes (\star Fp)}{(\star Fp)^{2}}$ & 
$X_{k}^{2} = 2\left(\dot{w}_{k} + 
2\mathscr{G}^{2}\ddot{w}_{k}\right)(\star 
Fy)^{2}$ \\
\hline
$\mathbf{P}_{3} = \frac{(Fp)\otimes (\star Fp)}{\mathscr{G}p^{2}}$ & $X_{k}^{3} 
= 2\mathscr{G}^{2}\dot{w}_{k}^{\prime}y^{2}$ \\
\hline
$\mathbf{P}_{4} = \frac{(\star Fp)\otimes (Fp)}{\mathscr{G}p^{2}}$ & $X_{k}^{4} 
= X_{k}^{3}$ \\
\hline
\end{tabular}
\caption{Algebraic expressions for projections and their respective coefficients according 
to the expansion in \Eqref{eq:c.e}. The symbol $\otimes$ denotes the dyadic 
product on $\mathbb{R}^{4}\times\mathbb{R}^{4}$. Here, $p$ is a dimensionful momentum space coordinate, and $y = p/k$ its dimensionless complement.}
\label{tab:c.a}
\end{table}

\noindent The class of projectors $\mathbf{P}_{a}$ for $a \in \lbrace 
1,2,3,4\rbrace$ refer to further subtransversal projections, because 
$\mathbf{P}_{a}\circ\mathbf{P}_{L} = \mathbf{P}_{L}\circ\mathbf{P}_{a} = 0$. 
Due to $\mathds{1} = \mathbf{P}_{T} + \mathbf{P}_{L}$, we have 
$\mathbf{P}_{a}(\mathfrak{X}^{*}(\mathbb{R}^{4})) \subseteq 
\mathbf{P}_{T}(\mathfrak{X}^{*}(\mathbb{R}^{4}))$ for all $a$.
From compositions among subtransversal
projections $\mathbf{P}_{a}$, it is further noted that they agree on
an equal rank and likewise share the \textit{same} one-dimensional image. Hence
it is not possible to span the three-dimensional transversal subspace
using a combination of the $\mathbf{P}_{a}$; however, we can
still make use of their properties evaluating the explicit form of
the flow and allowing us to estimate the weight of individual contributions to it.

Correspondingly, we span the regulator $\mathcal{R}_{k}$ with the aid of the
transversal and longitudinal projectors, yielding in momentum 
space:
\begin{equation}
\mathcal{R}_{k}(y) = 
Z_{k}k^{2}y^{2}\, r\left(y^{2}\right)\left[\mathbf{P}_{T} + 
\frac{1}{\alpha}\mathbf{P}_{L}\right],
\label{eq:c.h}
\end{equation}
where the information about the details of the momentum-mode regularization are
encoded in the dimensionless shape function $r$. Equations (\ref{eq:c.e}) and
(\ref{eq:c.h})  read together form the inverse of the regularized propagator,
$\Gamma_{k}^{(2)} + \mathcal{R}_{k} \equiv G_{k}^{-1}$, for which we now need the
operator inverse.  
For this, we take advantage of the algebraic structure provided by the field 
space projections and expand the regularized full propagator as in \Eqref{eq:c.e}:
\begin{equation}
\begin{split}
G_{k} &= \left(\Gamma_{k}^{(2)} + \mathcal{R}_{k}\right)^{-1}  \\
&= 
\frac{1}{Z_{k}k^{2}}\left(Y_{k}^{T}\mathbf{P}_{T} + Y_{k}^{L}\mathbf{P}_{L} + 
\sum_{a=1}^{4}Y_{k}^{a}\mathbf{P}_{a}\right).
\end{split}
\label{eq:c.i}
\end{equation}
The coefficients $Y_{k}^{T},Y_{k}^{L},Y_{k}^{a}$ for $a \in \lbrace 1,2,3,4\rbrace$ are 
completely determined by the system of equations that follows from $\mathds{1} = 
G_{k}^{-1}G_{k}$, using the composition table for the projections. 
The solution is
\begin{widetext}
\begin{equation}
\begin{split}
&\hspace*{4cm}Y_{k}^{T} = \frac{1}{X_{k}^{T} + y^{2}r}, \qquad Y_{k}^{L} = 
\frac{1}{X_{k}^{L} + \frac{1}{\alpha}y^{2}r}, \\[1em]
&Y_{k}^{1} = Y_{k}^{T}\cdot\frac{(X_{k}^{3})^{2}\xi^{-2} - X_{k}^{1}(X_{k}^{T} + X_{k}^{2} 
+ y^{2}r)}{(X_{k}^{T} + X_{k}^{1} + X_{k}^{3} + y^{2}r)(X_{k}^{T} + 
X_{k}^{2} + X_{k}^{3} + y^{2}r) - (X_{k}^{3} + X_{k}^{2}\xi^{2})(X_{k}^{1} + 
X_{k}^{3}\xi^{-2})},\\[1em]
&Y_{k}^{2} = Y_{k}^{T}\cdot\frac{(X_{k}^{3})^{2}\xi^{-2} - X_{k}^{2}(X_{k}^{T} + X_{k}^{1} 
+ y^{2}r)}{(X_{k}^{T} + X_{k}^{1} + X_{k}^{3} + y^{2}r)(X_{k}^{T} + 
X_{k}^{2} + X_{k}^{3} + y^{2}r) - (X_{k}^{3} + X_{k}^{2}\xi^{2})(X_{k}^{1} + 
X_{k}^{3}\xi^{-2})},\\[1em]
&Y_{k}^{3} = Y_{k}^{T}\cdot\frac{X_{k}^{1}X_{k}^{2}\xi^{2} - X_{k}^{3}(X_{k}^{T} + 
X_{k}^{3} + y^{2}r)}{(X_{k}^{T} + X_{k}^{1} + X_{k}^{3} + 
y^{2}r)(X_{k}^{T} + X_{k}^{2} + X_{k}^{3} + y^{2}r) - (X_{k}^{3} + 
X_{k}^{2}\xi^{2})(X_{k}^{1} + X_{k}^{3}\xi^{-2})}, \\[1em] 
&Y_{k}^{4} = Y_{k}^{3}, \\[1em]
&\hspace*{4.2cm}\text{with} \,\,\, \xi(F,y) := 
\frac{\mathscr{G}y^{2}}{\sqrt{(Fy)^{2}(\star Fy)^{2}}}.
\end{split}
\label{eq:c.ia}
\end{equation}
\end{widetext}

As a useful consequence of the projection technique, the RHS of the flow
equation \eqref{eq:b.j} decomposes into a sum of traces over field space
projectors.

Finally, the $\mathbf{Tr}$ operation in the flow equation runs over  
Lorentz indices and momentum space. Furthermore, because of field-strength 
homogeneity, the RG flow is projected onto the field-strength potential $w_{k}$. Introducing an RG 
time $t := \ln(k/\Lambda)$, with an arbitrary reference scale $\Lambda \in 
\mathbb{R}^{+}$, the RG flow finally is described by an autonomous differential equation that reads:
\begin{equation}
\begin{split}
&\partial_{t}w_{k} + 4w_{k} - (4 + \eta_{k})\left(w_{k}^{\prime}\mathscr{F} + 
2\dot{w}_{k}\mathscr{G}^{2}\right)\\[1em] &\hspace*{0.2cm}= 
-\frac{1}{32\pi^{4}}\int\limits_{\mathbb{R}^{4}}y^{2}\left(\eta_{k}r(y^{2}) + 
2y^{2}r^{\prime}(y^{2})\right){Y}_{k}(y)\,\mathrm{d}^{4}y,
\end{split}
\label{eq:c.k}
\end{equation}
where $r^{\prime}$ denotes the derivative of $r$ with respect to its argument $y^{2}$ and
\begin{equation}
{Y}_{k} := 3Y_{k}^{T} + \frac{1}{\alpha}Y_{k}^{L} + Y_{k}^{1} + Y_{k}^{2} 
+ 2Y_{k}^{3}.
\label{eq:c.l}
\end{equation}
Equation \eqref{eq:c.k} generalizes previously derived flow equations for
actions depending solely on $\mathscr{F}$ \cite{Gies:2004hy,Laporte:2021kyp} to
the general case of nonlinear ED, representing an important intermediate result
of this work. 

\subsection{Fixed-Point Sector}
\label{par:threethree}

In perturbative QED, the flow analogous to \Eqref{eq:c.k} develops singularities
toward high energies, e.g., in the form of the Landau pole. By contrast, the RG
flow can be UV complete if all operators spanning the action remain bounded. The
latter can be realized with the aid of RG fixed points where the dimensionless
flow vanishes and the theory develops a quantum scale symmetry
\cite{Wetterich:2019qzx}. The existence of such fixed points is a prerequisite
for the asymptotic-safety scenario of quantum field theories.

In the following, we address the question as to whether the quantized version of
nonlinear ED as described by the RG flow of \Eqref{eq:c.k} exhibits such a fixed
point.
If so, the scale derivative of the field-strength potential vanishes,
$\partial_{t}w_k=0$, and the potential approaches a \textit{fixed function}
$w_{*} := w_{k\to\infty}$; in the language of statistical mechanics, $w_\ast$
corresponds to a scaling function. A special focus on properties like nontriviality and
global existence for the
fixed function, if it exists, will be adopted later.

At the fixed point, $w_\ast$ satisfies the
fixed-function equation (FFE):
\begin{equation}
    \begin{split}
    &w_{*} - \left(1 + \frac{\eta_{*}}{4}\right)\left(w_{*}^{\prime}\mathscr{F} + 2\dot{w}_{*}\mathscr{G}^{2}\right)
    \\ &\hspace*{0.2cm}= - \frac{1}{128\pi^{4}}\int\limits_{\mathbb{R}^{4}}y^{2}\left(\eta_{*}r(y^{2}) + 2y^{2}r^{\prime}(y^{2})\right)Y_{*}(y)\,\mathrm{d}^{4}y.
    \end{split}
    \label{eq:c.m}
\end{equation}
Here and in the following, quantities evaluated at the fixed point are denoted
with an asterisk. In particular, the quantity $Y_{*}$ is obtained by evaluating \Eqref{eq:c.l}
for $w_k = w_\ast$; note that $Y_{*}$ contains derivatives of $w_{*}$ up to
second order in both arguments, such that \Eqref{eq:c.m} corresponds to a partial differential equation for $w_\ast$ as a function of $\mathscr{F}$ and $\mathscr{G}^2$.

\subsection{Approximations}
\label{par:threefour}

In order to investigate the fixed-function equation (\ref{eq:c.m}), we
use two approximations for simplifying the technical complexity:
\\[1em]
\indent \textbf{(A1)} The RHS of the FFE involves a momentum-space integral where
spherical symmetry is broken by the directions of electric and magnetic fields.
For instance, the term $(Fy)^{2}$ can be written as
\begin{equation}
\begin{split}
(Fy)^{2} &= y_{0}^{2}E^{2} + \vec{y}^{2}E^{2}\cos(\vartheta_{E})^{2}\\
&\hspace*{0.5cm} + \vec{y}^{2}B^{2}\sin(\vartheta_{B})^{2} + 2y_{0}\vec{y}
\cdot\left(E\times B\right),
\end{split}
\label{eq:c.n}
\end{equation}
where we have used a Euclidean space-time decomposition of $y =
(y_{0},\vec{y})^{\mathsf{T}}$ with $\vec{y} \in \mathbb{R}^{3}$, and
$\vartheta_{E},\vartheta_{B}$ denote the angles enclosed by $\vec{y}$ and $E,B$
respectively. The angle dependence can be eliminated by (\textit{i}) assuming
that $\vartheta_{E} - \vartheta_{B} = n\pi$ for $n \in \mathbb{Z}$, implying
that $E$ and $B$ are either parallel or antiparallel, and (\textit{ii})
requiring $E^{2} = B^{2}$. It can be shown, that these conditions are equivalent
to \textit{(anti-)self-dual nonlinear electrodynamics} for which $F =
\pm(\star F)$ (where the minus sign corresponds to anti-self-duality); note that self-duality as used here is  meaningful only in
$d = 4$ dimensions. Using self-duality, \Eqref{eq:c.n} exhibits spherical
symmetry in momentum space,
\begin{equation}
(Fy)^{2} = \mathscr{F}y^{2} = (\star Fy)^{2},
\label{eq:c.o}
\end{equation}
such that the momentum integral in \Eqref{eq:c.m} can be done analytically for
a variety of frequently used shape functions.

We emphasize that the choice of a self-dual field configuration for the
evaluation of the RHS of the FFE does not yet represent an approximation. The
RHS still contains the complete set of terms. With this choice, we, however,
lose the ability to distinguish between dependencies of the RHS on the two
different variables $\mathscr{F}$ or $\mathscr{G}$. In fact, since
$\mathscr{G}^{2}= \mathscr{F}^{2}$, every term of even power in $\mathscr{F}^{2}$ receives also contributions from the $\mathscr{G}^{2}$ dependence. Our first approximation \textbf{(A1)} therefore consists in accepting the (mis)identification of these terms on the RHS of the flow equation. 
As an advantage, the FFE now reduces to an ordinary differential equation as the
bi-argument dependence of the fixed function $w_{*}$ merges to a single-argument
dependence solely on the invariant $\mathscr{F}$.
\\[1em]
\indent \textbf{(A2)} The FFE can also be transformed into an ordinary differential
equation by truncating the theory space down to a pure $\mathscr{F}$ dependence
of $w_\ast$, i.e., discarding the $\mathscr{G}^{2}$ dependence altogether. Then,
the fact that $\dot{w}_{*} = \ddot{w}_{*} = \dot{w}_{*}^{\prime} = 0$ implies
that $Y_{k}^{2} = Y_{k}^{3} = Y_{k}^{4} = 0$ according to Table~\ref{tab:c.a} and
using (\ref{eq:c.ia}).

As a consequence, the scale-dependent propagator $G_{k}$ receives only a single
subtransversal contribution, cf.~\Eqref{eq:c.i}. Nevertheless, we believe that
this truncation still provides a good approximation, since the subtransversal
input arises from a unique one-dimensional subspace of $\mathfrak{X}^{*}(\mathbb{R}^{4})$ and
is likely to be of minor relevance compared to the full three-dimensional
transversal input generated by $\mathbf{P}_{T}$. The latter is mediated through
the coefficient $Y_{k}^{T}$ which remains unaffected from this truncation. At
the same time, this approximation yields a considerable simplification of
\Eqref{eq:c.m}.

Conversely, it would not be reasonable to restrict to a purely
$\mathscr{G}^{2}$-dependent theory space and discard the $\mathscr{F}$ sector
instead. This would eliminate the transversal input, only retaining
one-dimensional contributions that would not cover the underlying four-dimensional
field space. Also the Maxwell term would be discarded from the weak-field
expansion of $w_{*}$, thereby losing a relevant part of theory space including
the free theory and propagator.
\\[1em]
\indent We emphasize, that the approximations \textbf{(A1)} and \textbf{(A2)}
are not equivalent. Clearly \textbf{(A2)} $\nRightarrow$ \textbf{(A1)}, because
a truncation of theory space does not induce any specification of the field
strength used to build the invariants. \textbf{(A1)} is a restriction rather on
the information extracted for the fixed function $w_{*}$ than on the theory
space. In addition, also \textbf{(A1)} $\nRightarrow$ \textbf{(A2)}, since
self-duality retains information about the $\mathscr{G}^{2}$ dependence by means
of a projection on the $\mathscr{F}$-related subspace of theory space. In this
manner, derivatives of $w_{*}$ with respect to its $\mathscr{G}^{2}$ argument
transform to $\mathscr{F}$ derivatives and, in particular, do not imply
$\dot{w}_{*} = 0$. This is different from the demands of \textbf{(A2)}.

In previous studies \cite{Gies:2004hy,Laporte:2021kyp}, the truncation represented by
approximation \textbf{(A2)} has been applied. Specifically in
\cite{Laporte:2022ziz}, the problem of the angle dependence has been solved by
an expansion and resummation technique. As we apply both \textbf{(A1)} and
\textbf{(A2)}, our resulting FFE differs from that of \cite{Laporte:2022ziz} by
the terms kept from the identification of $\mathscr{G}^{2}=\mathscr{F}^{2}$. On
the other hand, by performing the momentum integration without expansion we have
an unaffected access to the large-field limit of the FFE.
\section{Fixed Functions for Improper Initial Conditions}
\label{sec:four}
In this section, we focus on the reduced FFE that we obtain based on the
approximations \textbf{(A1)} and \textbf{(A2)}. We search for solutions
employing a weak-field expansion technique that is widely used in the
literature, e.g., for the analysis of Wilson-Fisher-type fixed points in scalar
field theories in a local-potential approximation
\cite{Tetradis:1993ts,Morris:1998da,Berges:2000ew,Litim:2001up,Delamotte:2007pf,Litim:2016hlb,Zambelli:2016cbw},
generic fermionic or Yukawa models
\cite{Hofling:2002hj,Braun:2010tt,Braun:2011pp,Mesterhazy:2012ei,Jakovac:2014lqa,Gies:2013fua},
supersymmetric models
\cite{Synatschke:2008pv,Gies:2009az,Synatschke:2010ub,Heilmann:2012yf}, or
asymptotically safe fixed points in gravity
\cite{Codello:2006in,Machado:2007ea,Benedetti:2009rx,Falls:2013bv,Falls:2014tra,Eichhorn:2015bna,Kluth:2022vnq}. 

Even though this method is simple and has proven to lead to robust results in
many examples, it is based on a choice of \textit{improper} initial conditions
for the FFE that do not fix the solution uniquely without additional assumptions
and may, in fact, produce artifacts as will be discussed below. Indeed, our
results are similar to those of \cite{Laporte:2022ziz,deBrito:2023myf} as we
find many fixed-point candidates in addition to the Gaussian fixed point. Subsequently, we will, however, argue that \textit{proper} initial conditions give a more immediate access to fixed-point candidates for the present system.

\subsection{Reduced Fixed-Function Equation}
\label{par:fourone}

Applying both approximations \textbf{(A2)} \& \textbf{(A1)} in
this order reduces \Eqref{eq:c.m} significantly:
\begin{equation}
\begin{split}
&w_{*} - \left(1 + \frac{\eta_{*}}{4}\right)w_{*}^{\prime}\mathscr{F} = \frac{1}{64\pi^{2}}\biggl[3\mathrm{t}_{(1 \, 0)}^{4}\left(w_{*}^{\prime};
\frac{\eta_{*}}{2}\right) \\[0.5em] &\hspace*{0.3cm}+ \mathrm{t}_{(1 \, 0)}^{4}\left(1;\frac{\eta_{*}}{2}\right) - w_{*}^{\prime\prime}\mathscr{F}\mathrm{t}_{(1 \, 1)}^{4}
\left(w_{*}^{\prime},(w_{*}^{\prime}\mathscr{F})^{\prime};\frac{\eta_{*}}{2}\right)\biggr],
\end{split}
\label{eq:d.a}
\end{equation}
where we have introduced the \textit{threshold functions}
\begin{equation}
\begin{split}
&\mathrm{t}_{(n_{1} \, n_{2})}^{d}(z_{1},z_{2};a) \\[0.5em]
&\hspace*{1cm} := \int\limits_{0}^{\infty}y^{\frac{d}{2}-1}
\frac{-yr^{\prime}(y) - ar(y)}{\left(z_{1} + r(y)\right)^{n_{1}}\left(z_{2} + r(y)\right)^{n_{2}}}\,\mathrm{d}y.
\end{split}
\label{eq:d.b}
\end{equation}
If either $n_{1} = 0$ or $n_{2} = 0$, $\mathrm{t}$ does not depend on $z_{1}$
or $z_{2}$ respectively and we will just omit the redundant argument in our
notation, e.g., $\mathrm{t}_{(n_{1} \, 0)}^{d}(z_{1};a)$.

In the remainder of this section, we attempt to  construct a solution to \Eqref{eq:d.a} using analytical techniques based on small-field expansions.

\subsection{Small-Field Expansion}
\label{par:fourtwo}

Assuming that the field-strength potential at the fixed point can be expanded in a Taylor series
near the origin,
\begin{equation}
\begin{split}
w_{*}(\mathscr{F}) &
= \sum_{i=0}^{\infty}
u_{i,\ast}\mathscr{F}^{i} \\[0.5em] &\hspace*{-0.5cm}= u_{0,*} + \mathscr{F}
+ u_{2,*}\mathscr{F}^{2} + O(\mathscr{F}^{3}) \quad \text{as $\mathscr{F}
 \rightarrow 0$},
\end{split}
\label{eq:d.c}
\end{equation}
we need to determine the generalized fixed-point couplings $u_{i,*}$ from the FFE
\eqref{eq:d.a}.

We implement the IR regularization using the Litim-type regulator shape function \cite{Litim:2000ci},
\begin{equation}
r(y) = \frac{1-y}{y}\,\mathbf{1}_{[0,1)}(y),
\label{eq:d.d}
\end{equation}
in which $\mathbf{1}_{[0,1)}$ denotes the characteristic function on the
semi-open interval $[0,1) \subset \mathbb{R}$. This choice transforms all
threshold functions (\ref{eq:d.b}) occurring in  \Eqref{eq:d.a} into the following:
\begin{equation}
\begin{split}
&\mathrm{t}_{(1 \, 0)}^{4}\left(w_{*}^{\prime};\frac{\eta_{*}}{2}\right) = \frac{1}{2}\int\limits_{0}^{1}y\frac{2 - \eta_{*}(1-y)}{1-(1-w_{*}^{\prime})y}\,\mathrm{d}y,
\\[0.5em]
&\mathrm{t}_{(1 \, 0)}^{4}\left(1;\frac{\eta_{*}}{2}\right) = \frac{1}{2}\left(1 - \frac{\eta_{*}}{6}\right), \\[0.5em]
&\mathrm{t}_{(1 \, 1)}^{4}\left(w_{*}^{\prime},(w_{*}^{\prime}\mathscr{F})^{\prime};\frac{\eta_{*}}{2}\right) = \\[0.5em]
&\hspace*{0.2cm}\frac{1}{2}\int\limits_{0}^{1}y^{2}\frac{2 - \eta_{*}(1-y)}{\bigl(1 - 
(1-w_{*}^{\prime})y\bigr)\bigl(1-(1-(w_{*}^{\prime}\mathscr{F})^{\prime})y\bigr)}
\,\mathrm{d}y.
\end{split}
\label{eq:threshold}
\end{equation} \\
\indent Inserting the ansatz (\ref{eq:d.c}) into the reduced FFE (\ref{eq:d.a})
and expanding the RHS in powers of $\mathscr{F}$, we obtain a tower of equations
for the generalized fixed-point couplings $u_{i,*}$ by comparison of coefficients, the first four of which are listed below:
\begin{equation}
\begin{split}
    u_{0,\ast} &= \frac{6-\eta_{*}}{192\pi^{2}}, \\
    \eta_{*} &= \frac{8u_{2,\ast}}{48\pi^{2} + u_{2,\ast}}, \\
    u_{2,\ast} &= \frac{1}{2}(4+\eta_{*})u_{2,\ast} + \frac{3}{320\pi^{2}}(10-\eta_{*})u_{2,\ast}^{2}  \\  &\hspace*{0.1cm}- \frac{5}{512\pi^{2}}(8-\eta_{\ast})u_{3,\ast}, \\
    u_{3,\ast} &= -\frac{1}{48\pi^{2}}(12-\eta_{*})u_{2,*}^{3} + \frac{3}{4}(4+\eta_{*})u_{3,*} \\ 
   &\hspace*{0.1cm}+ \frac{3}{80\pi^{2}}(10-\eta_{*})u_{2,*}u_{3,*} - \frac{1}{64\pi^{2}}(8-\eta_{*})u_{4,*}.
    \end{split}
\label{eq:SMEtower}
\end{equation}
We observe that the vacuum energy $u_{0,\ast}$ is fully determined by the
anomalous dimension and completely decouples from the higher-order couplings.
Having fixed the wave function renormalization such that $u_{1,\ast}=1$, the
anomalous dimension is fully defined in terms of $u_{2,\ast}$, such that the
equations for the higher-order couplings can structurally be written as
$u_{i,\ast}=f_{i}(u_{2,\ast},\dots,u_{i+1,\ast})$ for $i\geq 2$. The function
$f_{i}$ here features a linear dependence on the highest-order coupling
$u_{i+1,*}$ involved.

This structure illustrates the role of initial conditions: since the FFE
is a second-order ODE, two initial conditions are needed for specifying a
solution. One initial condition is given by $u_{1,\ast}=1$ through our choice of
the wave function renormalization. A solution of the tower of equations, say up
to order $N \in \mathbb{N}_{0}$, now requires one further initial condition. In
principle, any coupling $u_{2,\ast}, \dots, u_{N+1,\ast}$ could be fixed for
this purpose. In order to construct a systematic expansion scheme, the standard
strategy is to fix $u_{N+1,\ast}$ to some value, typically $u_{N+1,\ast}=0$, and
then increase $N$ until some convergence criterion is met.

On the level of the FFE, this strategy corresponds to the initial
conditions $w'_\ast(0)=1$ and $w^{(N+1)}_{*}(0)=0$. We call these initial conditions
\textit{improper}, because ($i$) they do not guarantee a unique solution since the
conditions of the Picard-Lindelöf theorem are not matched, and ($ii$)
they do not cover the full space of possible initial conditions and thus of the
solution space.

\begin{figure}
    \begin{center}
    \includegraphics[scale=0.55]{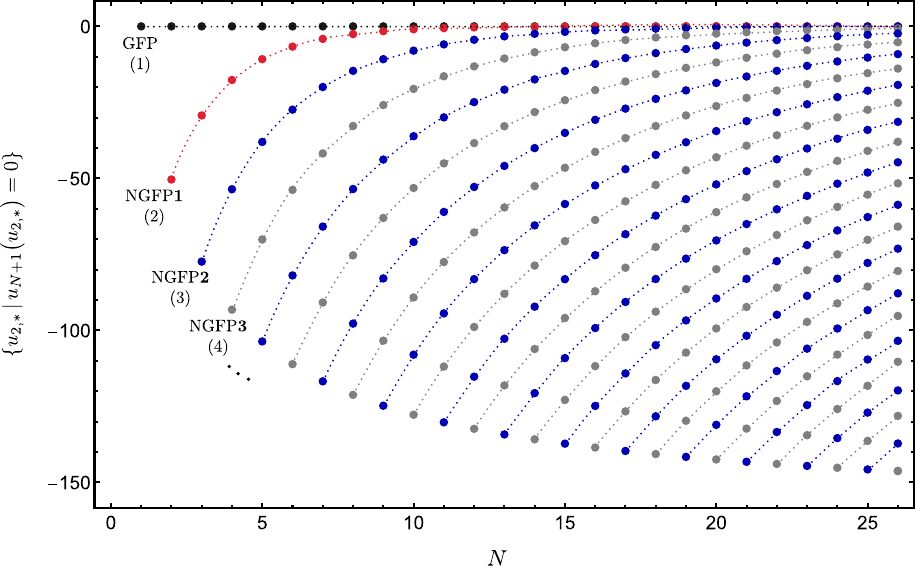}
    \caption{Fixed-point candidates within the small-field expansion using improper initial conditions, $u_{N+1,*}(u_{2,*}) = 0$, as a function of  truncation order $N= 1, \dots, 26$. The dotted lines connect fixed-point candidates with the same number of relevant directions as labeled in parentheses below the fixed-point designations: GFP, NGFP\textbf{1}, etc. In addition to the Gaussian fixed point (black), we observe a fixed-point candidate with one relevant direction (red) and subsequent higher-order candidates (blue and gray), each of which moves toward weaker fixed-point couplings for increasing truncation order.}
    \label{fig.d.a}
    \end{center}
\end{figure}

In order to make this more precise, we define the partial sum
$w_{*}(\mathscr{F};N) := \sum_{i=0}^{N}u_{i,*}\mathscr{F}^{i}$ as a truncation
at order $N \in \mathbb{N}_{0}$. We can solve the resulting tower of equations
\eqref{eq:SMEtower}, expressing all couplings as functions of $u_{2,\ast}$,
i.e., $u_{i,\ast}=u_{i,*}(u_{2,\ast})$ for all $i \in \mathbb{N}_{0,\leq N}$.
The improper initial condition $u_{N+1,\ast}(u_{2,\ast})=0$ yields a polynomial
equation of degree $2N-1$ in $u_{2,*}$ for $N \geq 1$. Therefore, it can at most
admit $2N-1$ zeroes. For each odd $N$, we find $N (\leq 2N-1)$ distinct real
solutions, whereas for each even $N$ we instead obtain $N+1 (\leq 2N-1)$
such solutions. The case $N=0$ corresponds to $\eta_{*}(u_{2,*}) = 0$
and immediately implies $u_{2,*} = 0$ which describes the trivial solution. The
solutions up to order $N=26$ are shown in Fig.~\ref{fig.d.a} (except for one
solution existing only for even $N$ which we consider as an artifact).

In addition to the noninteracting Gaussian fixed point (GFP) characterized by vanishing fixed-point couplings, we observe non-Gaussian fixed points (NGFP) which can be classified according to their number of RG relevant directions. The latter are characterized by the critical exponents $\Theta_j^{(N)}$ derived from the spectrum of the truncated stability matrix $B_\ast^{(N)}$, being the Jacobian of the (column) vector of beta functions,  $\beta^{(N)} = (\beta_{0},\beta_{2},\beta_{3},\ldots , \beta_{N})^{\mathsf{T}}$ where $\beta_{i} := \partial_{t}u_{i}(k)$ (note that $\beta_{1} = \partial_{t}u_{1}(k) = 0$), with respect to the (column) vector of couplings $u^{(N)} := (u_{0},u_{2},u_{3},\ldots , u_{N})^{\mathsf{T}}$ evaluated at the fixed-point candidate $u_{*}^{(N)}$,
\begin{equation}
B_{*}^{(N)} = (D\beta^{(N)})(u_{*}^{(N)}), \quad -\Theta_j^{(N)} \in \mathrm{eig} \big(B_\ast^{(N)}\big).
    \label{eq:stabmat}
\end{equation}
Positive $\Theta_j^{(N)}$ mark RG relevant directions, corresponding to
perturbations of a fixed point that grow large towards the IR and dominate the
long-range physics. The number of relevant directions near a fixed point
corresponds to the number of physical parameters to be fixed for predicting all
low-energy observables. The increasing number of fixed-point candidates for
increasing truncation order $N$ also exhibit an increasing number of relevant
directions. In Fig.~\ref{fig.d.a}, we have connected all fixed-point candidates
with the same number of relevant directions by dotted lines to guide the eye,
and labeled the non-Gaussian fixed points with the number of relevant directions
in parentheses. 

For the GFP, the only relevant direction is associated with the vacuum energy
$u_0$. For the interacting fixed points, further nontrivial relevant directions
appear in addition to that of the vacuum energy which remains exactly at
$\Theta_0^{(N)} \equiv \Theta_{0} =4$ for all $N \in \mathbb{N}_{0}$, reflecting its
canonical dimension. Classifying the non-Gaussian fixed-point
candidates according to the same number of relevant directions for increasing
truncation order in terms of fixed-point classes labeled by NGFP\textbf{n} as in
Fig.~\ref{fig.d.a}, we observe that each class is characterized by $n+1$
relevant directions, i.e. $n$ nontrivial directions apart from the one of the vacuum energy. 

If real, each of the NGFP\textbf{n} could represent a new universality class of
nonlinear electrodynamics giving rise to UV complete quantum field theories of
interacting light. However, following the standard reasoning in the literature
for small-field expansions, we expect only those NGFP candidates to approximate
a true interacting fixed point for which the couplings $u_{i,\ast}$ and the
critical exponents converge with increasing truncation order $N$.  

\begin{figure}
    \begin{center}
    \includegraphics[scale=0.55]{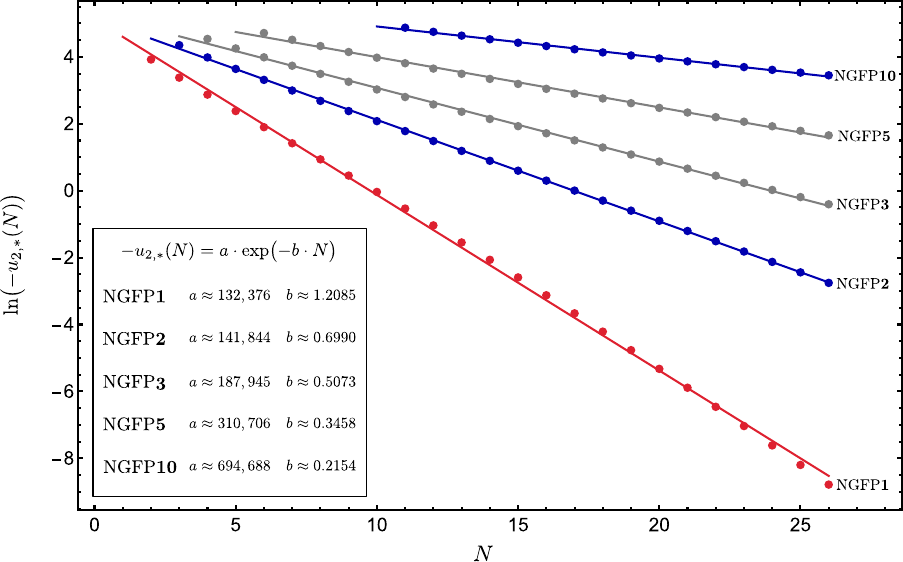}
    \caption{Logarithmic plot of the (negative) fixed-point coupling
    $-u_{2,*}(N)$ for selected interacting fixed point classes NGFP\textbf{n} for increasing truncation order $N$. The results from the truncated fixed-point equations can be fitted to an exponential with fit parameters $a,b$ such that $-u_{2,*}(N) = a\mathrm{e}^{-b N}$ (solid lines). The fit parameters are listed in the legend.}
    \label{fig.d.c}
    \end{center}
\end{figure}

From Fig.~\ref{fig.d.a}, all fixed-point candidates NGFP\textbf{n} appear to converge towards the GFP. In order to substantiate this quantitatively, we plot the fixed-point value $\ln(-u_{2,\ast})$ as a function of truncation order $N$ for a sample of the NGFPs in Fig.~\ref{fig.d.c}. We observe that the data follows linear fits, implying an exponential convergence with $N$: $u_{2,*}(N) =- a \exp(-bN)$ with fit constants $a,b$ for each non-Gaussian fixed-point class 
NGFP\textbf{n}, respectively.
Assuming that this exponential drop-off can be extrapolated to any $N$, we observe that $u_{2,\ast}$ and presumably also the higher-order couplings deplete to zero. Our data suggests that this also holds for higher-order NGFP\textbf{n} albeit with smaller damping rate. While this seems to suggest that all NGFP\textbf{n} converge to the GFP, we emphasize that we are dealing with different universality classes here, since the NGFP\textbf{n} exhibit a different number of relevant directions.

\begin{figure}
\begin{center}
\includegraphics[scale=0.55]{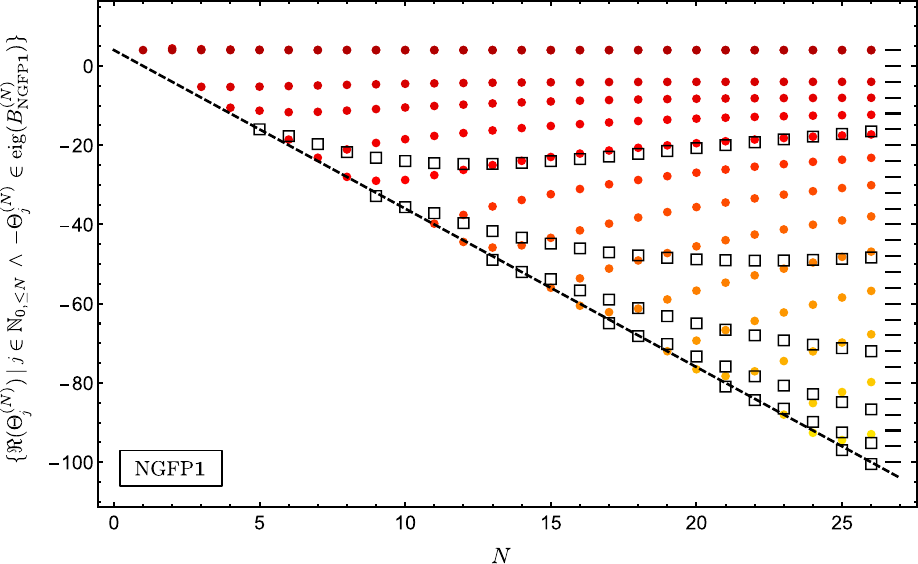}
\caption{Real parts ($\Re$) of critical exponents of the first non-Gaussian fixed point
(NGFP\textbf{1}) derived from the small-field expansion as a function of
truncation order $N$. Filled circles of equal color correspond to evolutions of
individual critical exponents for increasing $N$. Open squares mark the
real part of critical exponents which show up in complex conjugate pairs. The
black dashed line follows the canonical mass dimension of the highest
operator $\bar{\mathscr{F}}^{N}$ included at each $N$. Short solid ticks at the
right edge mark the (negative of the) canonical mass dimensions of all
operators occurring in the polynomial expansion, $4,0,-4,-8, \dots,-100$.}
\label{fig.d.d}
\end{center}
\end{figure}

In order to check for the convergence of the critical exponents, we concentrate
on  NGFP\textbf{1} (marked in red in Fig.~\ref{fig.d.a}) which exhibits one
nontrivial relevant direction in addition to the trivial vacuum-energy
direction. For this purpose, we track the evolution of
$B_{\mathrm{NGFP}\mathbf{1}}^{(N)}$ and its spectrum,
$\text{eig}(B_{\mathrm{NGFP}\mathbf{1}}^{(N)})$, for growing $N$. The result is
depicted in Fig.~\ref{fig.d.d}.
In addition to the critical exponent corresponding to $u_{0}$ which stays fixed
at $\Theta_{0}^{(N)} = 4$ for all $N\in\mathbb{N}_{0}$, the second relevant
exponent lies also close to the same value $\Theta_{2}^{(N)}\approx 4$; e.g., for
our highest
truncation, we find $\Theta_{2}^{(N=26)} - 4 < 10^{-5}$. The real parts of subsequent 
critical exponents remain negative (RG irrelevant) for all values of $N$ studied here,
and exhibit a clear tendency to approach the canonical mass dimensions of the
higher-order operators, cf. black ticks in Fig.~\ref{fig.d.d}, right side. In
particular, the convergence towards this asymptotic limit is already apparent
for the first few irrelevant exponents. In addition, we also find complex
conjugate pairs of critical exponents, the real parts of which are indicated by
open squares. In the $N$ range analyzed here, these complex pairs do not
yet exhibit a clear signature of convergence, as is also true for the
highest-order exponents. More definite answers would require higher
truncations.

Our findings so far show a close similarity to those of \cite{Laporte:2022ziz}
for nonlinear ED using a different truncation scheme as well as to those of
\cite{Laporte:2022ziz,deBrito:2023myf} studying shift-symmetric scalar field
theories both motivated by explorations of the weak-gravity bound
\cite{Eichhorn:2017eht,Eichhorn:2011pc,Eichhorn:2012va,Christiansen:2017gtg,deBrito:2021pyi,Knorr:2022ilz,Eichhorn:2019yzm,deBrito:2020dta,Eichhorn:2021qet}.
In fact, the resulting FFEs in these systems show a great deal of similarity
such that qualitative and even quantitative resemblance does not come as a
surprise.

In the light of this similarity, we expect that also the further results of
\cite{deBrito:2023myf} for the shift-symmetric scalar field are also of
relevance for nonlinear ED: for the eigenperturbations around the Gaussian
fixed point of the scalar theory, the corresponding differential equation can be
brought into Sturm-Liouville form which comes with an integration measure.
However, the eigenperturbations around fixed points analogous to our
NGFP\textbf{n} turn out not to be square-integrable with respect to the
Sturm-Liouville measure. Reference \cite{deBrito:2023myf} concludes for the shift-symmetric
scalar theory that these fixed points do not represent legitimate physical
fixed-point solutions and should be discarded, cf.
\cite{Halpern:1994vw,Halpern:1995vf,Morris:1994ki,Morris:1996nx,Gies:2000xr,Morris:2022rvd}.

Based on the strong similarity, we conjecture that an analogous Sturm-Liouville
analysis leads to the same verdict for the fixed points NGFP\textbf{n} derived
here from the small-field expansion using improper initial conditions.
Therefore, the only trustworthy fixed point so far is the trivial Gaussian one
with the field-strength potential $w_{\mathrm{GFP}}(\mathscr{F}) = \frac{1}{32\pi^{2}} + \mathscr{F}$.

\begin{figure}
\begin{center}
\includegraphics[scale=0.55]{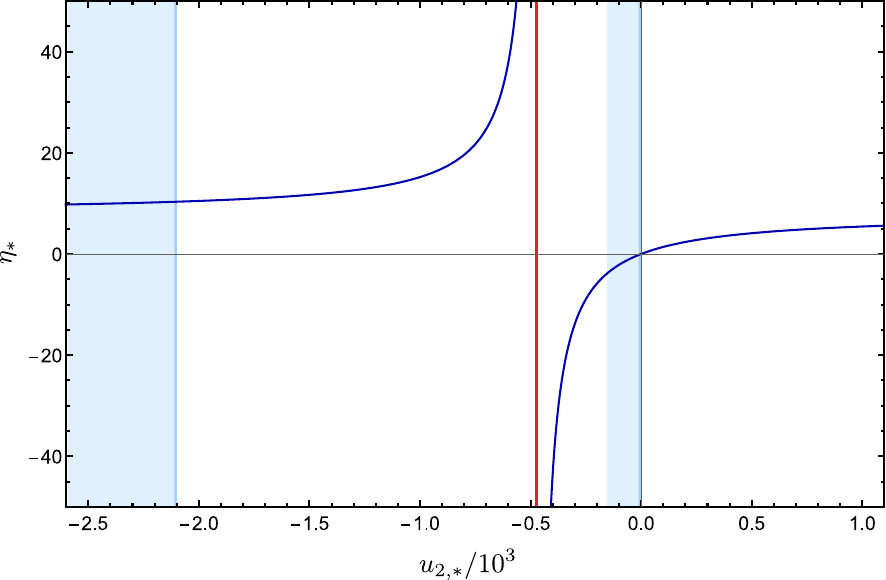}
\caption{Anomalous dimension $\eta_{*}$ plotted as a function of  the
coefficient $u_{2,*}$ of the $\mathscr{F}^{2}$ contribution to the small-field
expansion (blue line), exhibiting a pole at $u_{2,*} = -48\pi^{2}$ (red line).
The blue shaded areas indicate the regions where all fixed points from the classes
NGFP\textbf{n} based on the small-field expansion with improper initial
conditions for $N \in \mathbb{N}_{\leq 26}$ have been found; the blue shaded
segment at the left margin contains all fixed-point candidates classified as an artifact of the truncation, whereas the region right to the pole comprises the fixed-point candidates displayed in Fig.~\ref{fig.d.a}.}
\label{fig.d.b}
\end{center}
\end{figure}

This seems to suggest that no nontrivial fixed points exist in nonlinear
ED; however, the argument is incomplete: as we argue in the following, the
small-field expansion using improper initial conditions can be blind to further
solutions. In order to illustrate this already within the small-field
expansion, let us take a look at the behavior of the anomalous dimension
$\eta_{*}$ as a function of $u_{2,*}$. As discussed above, this relation is
fully determined by the $\mathscr{F}$-linear part of the reduced FFE and can be
found as the second equation from above in (\ref{eq:SMEtower}).
In fact, this relation also holds for more general Taylor expansions of
$w_{*}(\mathscr{F})$ around $\mathscr{F} = 0$, where $u_{2,*}$ has to be
replaced by $\frac{1}{2}w_{*}^{\prime\prime}(0)$. A plot of $\eta_\ast$ as a function of $u_{2,\ast}$ is shown in
Fig.~\ref{fig.d.b}.
For values $u_{2,*} \in (-\infty,-48\pi^{2})$, the anomalous dimension is positive
but large and reaches its limit point  $\eta_{*} = 8$ for
$u_{2,*} \rightarrow -\infty$. It moreover develops a pole at
$u_{2,*} = -48\pi^{2}$, where beyond that pole in the range
$u_{2,*} \in (-48\pi^{2},0]$ the anomalous dimension assumes negative
values and crosses the zero point for vanishing $u_{2,*}$. In the opposite half space of
positive $u_{2,*}$, $\eta_{*}$ is smooth throughout, slowly monotonically
growing and bounded from above by the limit value $\eta_{*} = 8$ for $u_{2,*}
\rightarrow \infty$. 

All non-Gaussian fixed points within the NGFP\textbf{n} displayed in Fig.
\ref{fig.d.a} are located in the slim blue shaded region between the Gaussian
solution and the pole. They cover thus a limited range of negative values for
$\eta_\ast$. The shifted branch of extra solutions existing only for
even $N$ and essentially ignored in the discussion above corresponds to the
separated blue shaded region at large $\vert u_{2,*}\vert$ beyond the pole. In
fact, classifying this shifted branch as an artifact is also justified by the
fact that the anomalous dimension is large, $\eta_\ast>8$. Since our ansatz for
the action is based on a derivative expansion, we expect the anomalous dimension
not to exceed values of $O(1)$ as a self-consistency criterion of the
expansion. 

Most importantly, we observe that the small-field expansion together with
the improper initial condition does not give access to solutions with small
positive values of $\eta_\ast$. From the viewpoint of \textit{proper initial
conditions}, this appears to be unnatural: a proper initial condition for the
FFE given, e.g., in terms of $w_{*}''(0) = u_{2,\ast}$ does naturally include small
positive values of $u_{2,\ast}$ implying likewise small positive values of
$\eta_\ast$.
Whether or not such initial conditions lead to a legitimate fixed point and a
global solution of the FFE needs to be and is studied separately in the
following sections.

Let us conclude this section with a few comments on the limitations of the
small-field expansion: In general, we expect the small-field expansion
(\ref{eq:d.c}) to have a finite radius of convergence (ROC). This radius
typically does not cover the maximal domain on which a full solution $w_{*}$ can
be defined, but rather a bounded interval $\mathscr{F} \in
\left[0,\mathscr{F}_{\mathrm{ROC}}\right)$. In the literature, numerical
shooting methods have frequently been used
\cite{Morris:1994ki,Codello:2012ec,Codello:2014yfa,Vacca:2015nta,Hellwig:2015woa}
to identify the initial condition for $u_{2,\ast}$ by that value that maximizes
$\mathscr{F}_{\mathrm{ROC}}$; this is based on the argument that a true
fixed-point solution should be globally defined. Since our current form of the
FFE is not suitable for shooting, we refrain from using this method, but
complement our approach by a large-field expansion below.

Finally, if we dropped assumption \textbf{(A2)} and reincluded
$\mathscr{G}^{2}$ dependencies into $w_{*}$, new features could appear in the
$\mathscr{F}$ part of theory space as the new couplings act nontrivially on the
RG flow. Since there is no concept of (Hodge) duality for a scalar field,
theories of nonlinear ED with fixed functions $(\mathscr{F},\mathscr{G}^{2})
\mapsto w_{*}(\mathscr{F},\mathscr{G}^{2})$ may no longer be comparable to
shift-symmetric scalar systems. This may in principle affect the implications
that we have conjectured from the FFE based on the analogy to the
shift-symmetric case. We will come back to this point in Sect. \ref{sec:six}.

\section{Fixed Functions for Proper Initial Conditions}
\label{sec:five}

In this section, we continue to use the approximations \textbf{(A1)} and
\textbf{(A2)}, but now aim at solving the FFE using \textit{proper initial conditions}:
as the reduced FFE (\ref{eq:d.a}) is a second order ordinary differential
equation, two initial conditions are required to single out a unique solution. As an example,
consider initial conditions at zero field amplitude, $w_{*}^{\prime}(0) = w_{1}$
and $w_{*}^{\prime\prime}(0) = w_{2}$ with constants $w_{1},w_{2}\in\mathbb{R}$.
As discussed above, $w_1=1$ is already fixed by our choice for the wave function
renormalization. This leaves us with a solution space, being a one-parameter
family $\lbrace \mathscr{F} \mapsto w_{*}(\mathscr{F} ; w_{2}) \, \vert \, w_{2}
\in \mathbb{R}\rbrace$.

In the present case where $w_{*}^{\prime\prime}(0) = u_{2,*}$, we could use
$w_{2} = u_{2,*}$ for parametrizing this family. For reasons that become clear
later, we use the inversion of the exact relation between $u_{2,\ast}$ and the
anomalous dimension $\eta_\ast$ in the second equation of (\ref{eq:SMEtower}), 
\begin{equation}
    u_{2,*}(\eta_{*}) = 48\pi^{2}\frac{\eta_{*}}{8 - \eta_{*}} \quad \text{for $\eta_{*} \neq 8$},
    \label{eq:e.a}
\end{equation}
in order to navigate through the space of solutions by using $\eta_{*}\in \mathbb{R}$ excluding the value $\eta_{*} = 8$ where (\ref{eq:e.a}) diverges.  

In order to single out the physical fixed-point solutions, another criterion is
needed. In many models, this criterion is given by global existence. For
instance, in Ising-type systems, a generic choice for $w_2$ yields a solution
with a singularity at a finite field amplitude
\cite{Hasenfratz:1985dm,Morris:1998da}, and only a single value of $w_2$ (or a discrete
set) corresponds to a solution which exists for any value of the amplitude. 

The following subsections are devoted to a construction of such global solutions
using the analytical tools of small- and large-field expansions.

\subsection{Small-Field Expansion}
\label{par:fivetwo}
Let us start again with the small-field expansion, now implementing proper
initial conditions. For this, we use again the Taylor expansion \eqref{eq:d.c},
leading to the tower of Eqs. \eqref{eq:SMEtower}. The essential difference
for a given value of $\eta_\ast$ now is that we can use the $i$th equation
$u_{i,\ast}=f_{i}(u_{2,\ast},\dots,u_{i+1,\ast})$ and solve it exactly for
$u_{i+1,\ast}$. Here we note two aspects: firstly, the solution is unique, since $f_i$ depends linearly on
$u_{i+1,\ast}$ and secondly, it is stable against increments of $N$ for every
admissible value of $\eta_{*}$. The latter means that the functional dependence
of $u_{i+1,*}$ on the anomalous dimension is unaffected from the order of
truncation and, once determined explicitly, applies to arbitrary $N$ (provided
that $i \leq N$, otherwise $u_{i+1,*}$ does not yet exist).
The explicit expressions for the first few coefficients
$u_{i,*}$ including the vacuum energy $u_{0,*}$ read:
\begin{equation}
\begin{split}
u_{0,*} &= \frac{6-\eta_{*}}{192\pi^{2}} , \\[0.5em]
u_{1,*} &= 1, \\[0.5em]
u_{2,*} &= 48\pi^{2}\frac{\eta_{*}}{8-\eta_{*}}, \\[0.5em]
u_{3,*} &= \frac{6144\pi^{4}}{25}\frac{\eta_{*}}{(8-\eta_{*})^{3}}\left(160+150\eta_{*}-19\eta_{*}^{2}\right) , \\[0.5em]
u_{4,*} &= \frac{49152\pi^{6}}{125}\frac{\eta_{*}}{(8-\eta_{*})^{5}}\left(102400+236800\eta_{*} \right. \\[0.5em] &\hspace*{1.6cm} \left. +67520\eta_{*}^{2}
-24520\eta_{*}^{3}+1563\eta_{*}^{4}\right) .
\end{split}
\label{eq:e.d}
\end{equation}
From investigating also higher order couplings we can find a general pattern,
according to which the $i$th coupling for $i \geq 2$ can be written as a function of $\eta_{*}$ as
\begin{equation}
    u_{i,*}(\eta_{*}) = A_{i}\pi^{2(i-1)}\frac{\eta_{*}}{(8-\eta_{*})^{2i-3}}P_{2(i-2)}(\eta_{*}),
    \label{eq:uietagenform}
\end{equation}
where $A_{i}$ is a number and $P_{D}$ denotes a full polynomial of
degree $D$ in $\eta_{*}$.

For illustration, let us study some explicit results for various choices of
$\eta_\ast$. Since the improper initial conditions gave us access to
negative values of $\eta_\ast$ only which we argued to correspond to artifacts
of the approximation, we now concentrate on the branch $\eta_\ast >0$.
Indeed, the fact that we are now capable of inspecting fixed points at
positive anomalous dimensions is a notable difference between improper and
proper initial conditions. As our truncation corresponds to a derivative
expansion, we expect our approximation to be justified for small values of
$\eta_{*} \lesssim O(1)$.
 
\begin{figure*}
\begin{center}
\includegraphics[scale=0.55]{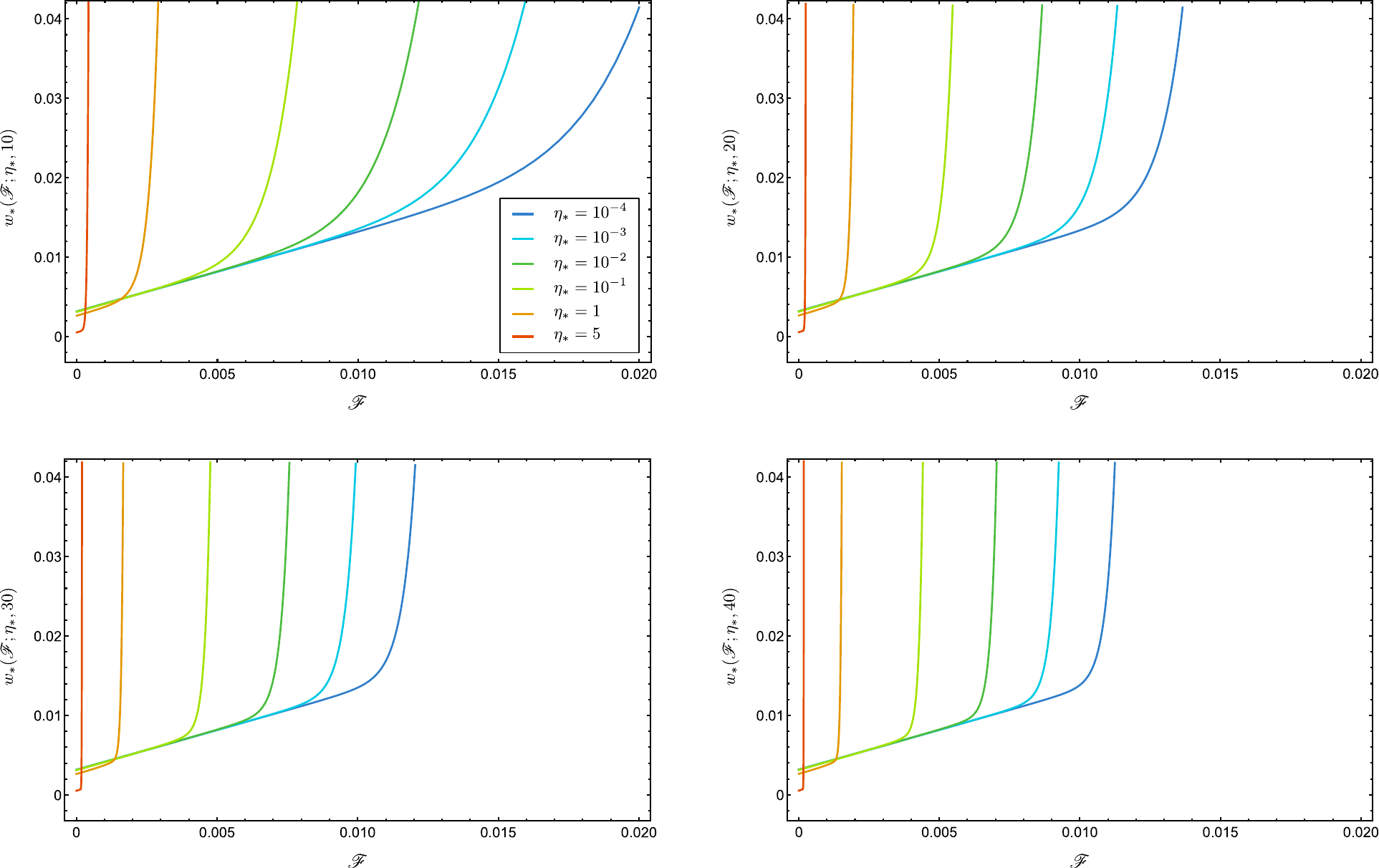}
\caption{(Partial) Fixed function $w_{*}$ plotted as a function of positive
$\mathscr{F}$ for different finite-dimensional truncations ($N$) and a selection
of $\eta_{*}$ values. The panes display results for increasing truncation order
from $N=10$ (upper left) to $N=40$ (lower right) in steps of $\Delta N = 10$. In
each panel, $w_{*}$ is shown for values of $\eta_{*}$ ranging logarithmically
from $10^{-4}$ to $1$ in colors from blue (right-most) to orange, respectively,
also including the extreme example $\eta_{*} = 5$ (red/left-most) for
illustrative purposes.}
\label{fig.e.a}
\end{center}
\end{figure*}

Let  $w_{*}(\mathscr{F};\eta_{*},N)$ denote the $N$th partial sum of
\Eqref{eq:d.c}. Several resulting field-strength potentials $\mathscr{F} \mapsto w_{*}(\mathscr{F};\eta_{*},N)$ in the range $\mathscr{F} \in [0,0.02]$,
for the choices $\eta_{*} \in \lbrace 10^{-4},
10^{-3},10^{-2},10^{-1},1,5\rbrace$ and $N \in \lbrace 10,20,30,40\rbrace$  are
shown in Fig.~\ref{fig.e.a}.
We observe that $w_{*}$ describes a monotonically increasing function with a
linear domain close to the origin, where approximately
$w_{*}(\mathscr{F};\eta_{*},N) -u_{0,*}(\eta_{*})\approx \mathscr{F}$. The range
of this domain depends sensitively on the two parameters $\eta_{*}$ and $N$. For
example, for increasing $N$ at fixed $\eta_{*}$, the point of departure from the
linear behavior is shifted to smaller $\mathscr{F}$. However, the speed of this
shift slows down rapidly for increasing $N$. Also for increasing $\eta_{*}$ at
fixed $N$, we observe a similar, if not more pronounced, effect.

Analogously to many other FRG studies of fixed-point potentials, we indeed
expect the small field expansion to exhibit a finite radius of convergence; the
preceding observed behavior is indicative for this. More quantitatively, let
$r(\eta_{*})$ denote the radius of convergence of the full power series
(\ref{eq:d.c}), our observations suggest that $r(\eta_{*})$ shrinks with increasing
$\eta_{*}$. From the polynomial relation $u_{i,*}(\eta_{*}) \propto P_{2i-4}(\eta_{*})$,
it is clear that $\lim_{\eta_{*}\rightarrow \infty}r(\eta_{*}) = 0$, whereas $r(0) = \infty$,
since $\eta_\ast=0$ corresponds to the Gaussian fixed point, where all $u_{i\geq 2,*}$
vanish identically.

\begin{figure}
\begin{center}
\includegraphics[scale=0.55]{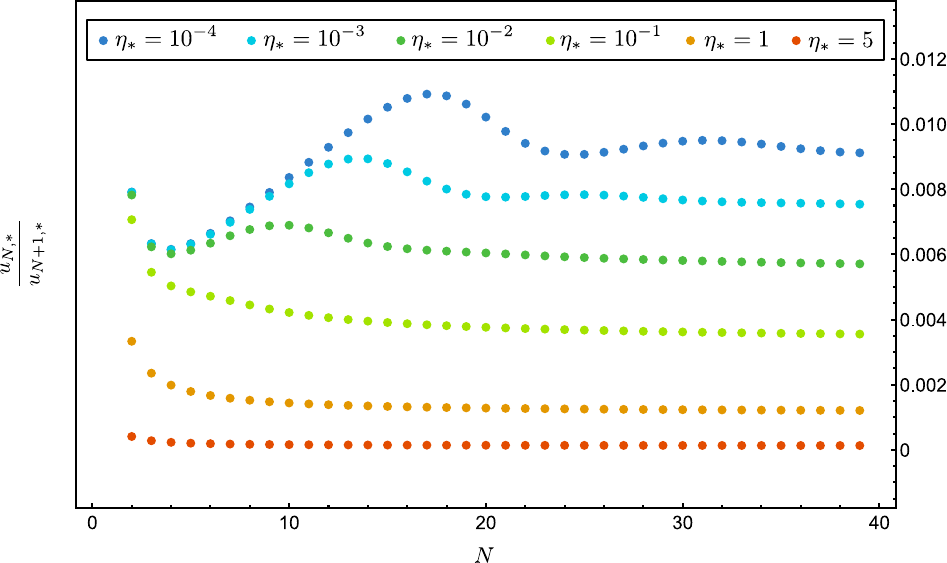}
\caption{Ratio sequence for the highest-order series coefficient $u_{N,*}$ of an
$N$th partial sum truncation of the small-field expansion \Eqref{eq:d.c}. Here, the set of $\eta_{*}$ values agrees with those of Fig.~\ref{fig.e.a}
following the same color code. }
\label{fig.e.b}
\end{center}
\end{figure}

In order to compute the radius of convergence in full generality we would need to use
Cauchy-Hadamard's theorem. Here, we confine ourselves to use a special case of the
theorem where $r$ can be extracted from the ratio test if all couplings are known,
{i.e.},
$r(\eta_{*}) = \lim_{i\rightarrow \infty}\vert u_{i,*}(\eta_{*}) / u_{i+1,*}(\eta_{*}) \vert$
whenever the limit exists, provided that all coefficients $u_{i,*}$ do not vanish above a
certain index. The latter requirement is certainly fulfilled in the case at hand and we
find the general result of the form
\begin{equation}
\frac{u_{i,*}(\eta_{*})}{u_{i+1,*}(\eta_{*})} = \frac{1}{\pi^{2}}\frac{A_{i}}{A_{i+1}}(8-\eta_{*})^{2}\frac{P_{2(i-2)}(\eta_{*})}{P_{2(i-1)}(\eta_{*})},
\label{eq:e.e}
\end{equation} 
in which $\frac{A_{i}}{A_{i+1}} < 1$ holds for all $i$ studied in this work. The
sequence (\ref{eq:e.e}) is depicted in Fig.~\ref{fig.e.b} for the values of
$\eta_{*}$ also considered in Fig.~\ref{fig.e.a}. Since each of these sequences
exhibits rapid convergence, we obtain estimates for $r(\eta_{*})$,
e.g., $r(10^{-4}) \approx 0.009$ or $r(1) \approx 0.001$ (and, as anticipated, $r(5)\approx 0$ for the extreme example of a large anomalous dimension $\eta_\ast=5$).
\subsection{Large-Field Expansion}
\label{par:fivethree}

On our way to construct global solutions to the FFE, let us next study a
large-field expansion. In order to identify a starting point, we use the
following line of argument: On the one hand, we expect the field-strength
potential to be an increasing function of the field amplitude. More
specifically, we expect $w_\ast(\mathscr{F})$ to diverge for
$\mathscr{F}\to\infty$, reflecting the fact that an infinitely large amplitude
should cost an infinite amount of Euclidean action. On the other hand, the
Sturm-Liouville analysis mentioned above suggests that the field-strength potential
should be polynomially bounded. It is thus natural to assume that the
field-strength potential diverges like a power for large amplitude,
$w_\ast(\mathscr{F})\sim \mathscr{F}^\Delta$ for
$\mathscr{F}\to\infty$, with a positive exponent $\Delta > 0$. If so, both terms on the left-hand side of
\Eqref{eq:d.a} scale like $\sim \mathscr{F}^\Delta$ to infinity at large fields.
On the right-hand side, we observe that all terms are bounded: for $\Delta>1$,
all field-dependent terms are suppressed $\sim 1/\mathscr{F}^{\Delta-1}$; for
$0<\Delta\leq 1$, the right-hand side approaches a constant.

Ignoring subleading constants, the reduced FFE (\ref{eq:d.a}) therefore
takes the asymptotic form
\begin{equation}
w_{*}  - \left(1 + \frac{\eta_{*}}{4}\right)w_{*}^{\prime}\mathscr{F} \, 
\sim \, 0 \quad (\mathscr{F} \rightarrow \infty).
\label{eq:e.i}
\end{equation}
Equation (\ref{eq:e.i}) corresponds to a first-order ordinary differential
equation which can easily be solved analytically, yielding a two-parameter
family of solutions with one integration parameter $\mu$ in addition to the
anomalous dimension $\eta_{*}$ parametrizing the proper initial condition of the
full equation:
\begin{equation}
\begin{split}
w_{*}(\mathscr{F};\mu,\eta_{*}) &\sim \mu\mathscr{F}^{\Delta(\eta_{*})} \quad 
(\mathscr{F} \rightarrow \infty), \\[0.5em]
&\hspace*{-0.4cm}\Delta(\eta_{*}) := \frac{4}{4+\eta_{*}}.
\end{split}
\label{eq:e.j}
\end{equation}
This demonstrates that our assumption of a power-law ansatz for the
large-field asymptotics is self-consistent as long as $\eta_\ast>-4$. Moreover,
we observe that this asymptotics is governed by the anomalous dimension. This is
in complete analogy to many other examples in the literature where the
large-amplitude asymptotics is balanced by the classical rescaling terms (i.e.,
the second term in \Eqref{eq:e.i}). For the construction of the global solution
below, the parameter $\mu$ will be fixed by the requirement of merging the
small- and large-field solutions.

Using \Eqref{eq:e.j} as a leading order, we now need an ansatz for a systematic
large-field expansion. For better readability, let us define
$\mathscr{X} := 1/ \mathscr{F}$ and rewrite the field-strength potential in terms of
$\mathscr{X}$, $\tilde{w}_{*}( \mathscr{X};\mu,\Delta(\eta_\ast)) =
w_\ast(\mathscr{F};\mu, \eta_\ast)|_{\mathscr{F}=\mathscr{X}^{-1}}$. Then, we may
parametrize $\tilde{w}_\ast$ for small $\mathscr{X}$ as 
\begin{equation}
\tilde{w}_{*}(\mathscr{X};\mu,\Delta) = c(\mu, \Delta) + \mu\mathscr{X}^{-\Delta} + \tilde{\mathcal{C}}_\ast(\mathscr{X};\Delta),
\label{eq:wtildeansatz}
\end{equation}
where $c$ is a constant for a fixed pair $(\mu,\Delta)$, and
$\tilde{\mathcal{C}}_\ast$ provides the subleading terms of higher orders
in $\mathscr{X}$. It is straightforward to check, that a naive
power-series ansatz for $\tilde{\mathcal{C}}_\ast$ in general leads to
artificial divergencies upon Taylor-expanding the threshold functions. Therefore
a more refined strategy is needed.

The field-dependent part of the integrands of the involved threshold functions given
in (\ref{eq:threshold}) can be written as either
\begin{equation}
\frac{1}{1-Ay} \qquad \text{or} \qquad \frac{1}{1-Ay}\cdot\frac{1}{1-By},
\label{eq:e.k}
\end{equation}
where $A$ and $B$ are $\mathscr{X}$-dependent quantities and $y \in [0,1]$ is
the variable of integration. The expressions in (\ref{eq:e.k}) can be expanded
into a geometric series as long as $\vert A\vert , \vert B\vert <1$. Given the ansatz
\eqref{eq:wtildeansatz} for $\tilde{w}_{*}$, these last-mentioned conditions can
be viewed as restrictive boundary conditions on the applicability domain of
$\mathscr{X}$, the range of which will be $\mu$ and $\Delta$ dependent. Using the
explicit forms of $A,B$ in \Eqref{eq:threshold}, these conditions read
\begin{equation}
\begin{split}
&0 < \Delta\mu\mathscr{X}^{1-\Delta}
-\mathscr{X}^{2}\tilde{\mathcal{C}}_{*}^{\prime}(\mathscr{X};\Delta) < 2,
\\[0.5em] &0 < \Delta^{2}\mu\mathscr{X}^{1-\Delta}+\mathscr{X}^{2}\Bigl(1 +
\mathscr{X}\tilde{\mathcal{C}}_{*}^{\prime\prime}(\mathscr{X};\Delta)\Bigr) < 2.
\end{split}
\label{eq:e.l}
\end{equation}
Here, a prime denotes the derivative with respect to the argument, e.g.,
$\tilde{\mathcal{C}}_{*}^{\prime}(\mathscr{X};\Delta) \equiv
\frac{\mathrm{d}\tilde{\mathcal{C}}_{*}(\mathscr{X};\Delta)}{\mathrm{d}\mathscr{X}}$.
Since $\tilde{\mathcal{C}}_{*}$ decreases fast enough for $\mathscr{X}
\rightarrow 0$ by assumption, these conditions can be fulfilled if $1-\Delta >
0$, such that contributions proportional to $\mathscr{X}^{1-\Delta}$ do not
become arbitrarily large. It is interesting to note that this implies, in particular,
$\eta_{*} > 0$, which is in fact the regime of interest to us.

The geometric series expansions of (\ref{eq:e.k}) on the one hand produce simple
integrals over positive powers of $y$ that can be performed analytically at any
order. On the other hand, the resulting field dependencies arise from the
corresponding powers $A^{n}$ or $B^{n}$ ($n\in\mathbb{N}_{0}$) which also
contain powers of $\tilde{w}_{*}$ and its derivatives. This produces monomials
$\propto \mathscr{X}^{m\Delta}$ ($m\in\mathbb{N}$) which cannot be covered by an
ordinary power-series ansatz for $\tilde{\mathcal{C}}_{*}$. It rather needs to
be expressed in terms of a formal Hahn series with $\Gamma(\Delta) \subset
\mathbb{N}^{2}$ being a suitable $\Delta$-dependent ordered group from which
the running index of the corresponding sum is taken. Our final
large-field ansatz results from an iterative process that covers all powers of
$\mathscr{X}$ arising from the geometric-series expansion:
\begin{equation}
\begin{split}
\tilde{w}_{*}(\mathscr{X};\mu,\Delta) &= 
\sum_{e\in\Gamma(\Delta)}v_{e}(\mu,\Delta)\mathscr{X}^{p(e)}\\[0.5em]
&\hspace*{-1.5cm}= c(\mu, \Delta) + \mu\mathscr{X}^{-\Delta} + 
\sum_{I=1}^{\infty}\sum_{a=1}^{I}v_{I}^{a}(\mu,\Delta)\mathscr{X}^{I-a\Delta}.
 \end{split}
\label{eq:e.m}
\end{equation}
Here, $v_{I}^{a}$ ($(I,a) = e \in \Gamma(\Delta) \subset \mathbb{N}^{2}$) are the $\mu$- and
$\Delta$-dependent coefficients to be determined and $p(e) := I - a\Delta$.

Unlike in the small-field regime, a truncation for the large-field sector will
thus be specified by two parameters $N_{1},N_{2} \in \mathbb{N}_{0}$. The former
truncates the first series (\ref{eq:e.m}) to its first $N_{1}$ terms
$I\in\mathbb{N}_{\leq N_{1}}$, including a total of
$\frac{1}{2}N_{1}(N_{1}+1)$ terms because of the double sum. Besides the
lowest order contribution, which is always $\mathscr{X}^{1-\Delta}$, this also
defines
the highest power of this expansion which is given by
$\mathscr{X}^{N_{1}-\Delta}$. Higher order contributions are neglected whenever
they are generated by the expansion of the reduced FFE. However, this procedure
is not fully self-consistent insofar as also contributions proportional to $\mathscr{X}^{p(e)}$
with $1-\Delta < p(e) < N_{1}-\Delta$ will emerge, but which are not part of our
truncation itself. Since those $p$'s can still be expressed as $p(e) = I-a\Delta$ for some pair $(I,a) = e \in \Gamma(\Delta)$, these
terms eventually get successively and consistently resolved at higher truncations $N_{1}$. In order to estimate the quantitative impact of these terms,
we will compare two different truncations and try to evaluate the weight of these
additional contributions. The second parameter $N_{2}$ limits the geometric
series emerging through (\ref{eq:e.k}) to contain only their first $N_{2}$
terms. The effect of choosing different $N_{2}$ values can be understood as
follows: if we think of all possible contributions $\propto \mathscr{X}^{I-a\Delta}$
being classified by an ordered sequence of sets which contain all exponents
according to their numerical value in the interval $I_{n} := [n,n+1)$ for
$n\in \mathbb{N}_{0}$, the parameter $N_{2}$ causes the number of terms
which actually appear on the RHS of the reduced FFE from each of these classes
to increase. For instance, if $\Delta$ is less than, but close to, one (which corresponds
to small $\eta_{*}$), terms with exponents $I-a\Delta$ for $a = I$ give
$I(1-\Delta) \in [1,2) = I_{1}$ for many values of $I$. Given a fixed parameter
value of $N_{1}$, we may generate more and more terms that belong to the class
$I_{1}$ which are yet not part of the truncation at hand if $N_{2}$ gets large
enough. In this sense, $N_{2}$ controls the resolution at which the spectrum of
all potential contributions taken from the classes $I_{n}$ and emerging on the
RHS of the reduced FFE is sampled. However, only if we also raise $N_{1}$ we
would be able to balance this effect with the LHS and reveal the information
available at this level of the resolution. Thus, it is advisable to narrow the
pertinent truncations on similar values, e.g., $N_{1} = N_{2} + 1$.
\subsection{Global Fixed Functions}
\label{par:fivefour}

Let us now proceed with the construction of global fixed functions based
on the analytic expansions for small and large fields employing proper initial
conditions. Since both expansions generically have a finite radius of
convergence, it is \textit{a priori} unclear whether both expansions have a finite
overlap region where they can be matched using the parameter $\mu$. For
instance, for scalar O($N$) models, such an overlap region does exist. If so, it
is typically not possible to perform the matching for any set of proper initial
condition parameters. In fact for scalar models this is possible only for a
discrete set of initial conditions that correspond to a discrete set of fixed
points \cite{Bridle:2013sra}, such as the Gaussian or the Wilson-Fisher fixed
point. If this standard scenario applies to the present case, we should expect
that it singles out specific values of $\eta_\ast$ for which global fixed
functions can be constructed. Incidentally, there is \textit{a priori} no guarantee that
a finite overlap region for the two expansions exists; see \cite{Schreyer:2020}
for a counter example. In this case, the present approach would not find a viable global solution and more powerful  methods such as those of \cite{Borchardt:2015rxa,Schreyer:2020,Grossi:2019urj,Koenigstein:2021syz} are needed.

Let us now construct estimates for global fixed functions obeying proper initial
conditions using the following steps:

\textbf{(1)} First, we construct a solution $w_{*}^{\mathrm{L}}$ from the
large-field expansion of the reduced FFE defined in terms of the two truncation
parameters $N_{1}\in\mathbb{N},N_{2}\in\mathbb{N}_0$. The ansatz (\ref{eq:e.m})
together with the reduced FFE yields an algebraic system from which we can
determine the constant $c$ as well as the $\frac{1}{2}N_{1}(N_{1}+1)$ unknown
coefficients $v_{I}^{a} \in \lbrace
v_{1}^{1},v_{2}^{1},v_{2}^{2},v_{3}^{1},\ldots ,v_{N_{1}}^{N_{1}}\rbrace$. The
latter are derived as functions of $\mu$ and $\Delta$.

\textbf{(2)} Second, we construct the small-field expansion $w_{*}^{\mathrm{S}}$
for proper initial conditions specified in terms of a value for $\eta_{*}$ based
on the highest-order truncation, i.e., the largest value of $N$ used in Sect.
\ref{par:fivetwo}. The proper initial condition, of course also fixes the
$\Delta=\Delta(\eta_\ast)$ dependence of the coefficients $v_{I}^{a}$, and thus
that of the large-field solution $w_{*}^{\mathrm{L}}$ from step \textbf{(1)}
which retains only a $\mu$ dependence.

\textbf{(3)} In order to specify the remaining free parameter $\mu$, let us
first quantify the overlap region of the two approximate solutions
$w_{*}^{\mathrm{S}}$ and  $w_{*}^{\mathrm{L}}$: from our construction of
$w_{*}^{\mathrm{S}}$, we also obtain an approximation of the radius of
convergence $r$, cf. \Eqref{eq:e.e} and Fig.~\ref{fig.e.b}. Unfortunately, a
formal Hahn series like (\ref{eq:e.m}) does generally not allow for a proper
notion of convergence, let alone a radius of convergence; still, we observe that
$w_{*}^{\mathrm{L}}$ develops a pronounced barrier $b_{\mathrm{L}}$ below
which the derivative $(w_{*}^{\mathrm{L}})^{\prime}$ rapidly increases. This
happens for sufficiently small $\mathscr{F}$ numerically comparable to some
power of the coefficients $v_{I}^{a}$. A numerical value for $b_{\mathrm{L}}$
can be estimated by $b_{\mathrm{L}} \approx \mathrm{max}\lbrace
(v_{I}^{a})^{\frac{1}{I-a\Delta}} \, \vert \, I \in \mathbb{N}_{\leq N_{1}}, a
\leq I\rbrace$. We use $b_{\mathrm{L}}$  as a provisional substitute for a
radius of convergence for $w_{*}^{\mathrm{L}}$.

Then, we define the overlap region of $w_{*}^{\mathrm{S}}$ and
$w_{*}^{\mathrm{L}}$ in terms of the interval intersection $[0,r]\cap
[b_{\mathrm{L}},\infty) = [b_{\mathrm{L}},r]$ if $b_{\mathrm{L}} \leq r$. (If
$b_{\mathrm{L}} > r$, there is no overlap region and a construction of a global
solution cannot be based on the small- and large-field expansions alone.) Even
though $b_{\mathrm{L}}$ carries a $\mu$ dependence in principle, we observe this
dependence to be rather weak; in practice, the approximation $b_{\mathrm{L}}
\approx \mathrm{const}.$ can hence be used for a wide range of $\mu$ values.

Now, we fix the free parameter $\mu$ by demanding that the square-deviation
integral in the overlap region,
\begin{equation}
\delta^{2}(\mu) := \int\limits_{[b_{\mathrm{L}},r]}\biggl(w_{*}^{\mathrm{S}}
(\mathscr{F}) - w_{*}^{\mathrm{L}}(\mathscr{F};\mu)\biggr)^{2}\,\mathrm{d}
\mathscr{F},
\label{eq:e.n}
\end{equation}
becomes minimal,
\begin{equation}
\mu \rightarrow \mu_{0}: \quad (\delta^{2})^{\prime}(\mu_{0}) = 0 \quad \land
\quad (\delta^{2})^{\prime\prime}(\mu_{0}) > 0.
\label{eq:e.o}
\end{equation}
In case of several local minima, we pick the $\mu_0$ value for the global
minimum of $\delta^{2}$. Inserting $\mu_{0}$ into $w_{*}^{\mathrm{L}}$ finally
completes the large-field solution.

\begin{figure*}[t]
\begin{center}
\includegraphics[scale=0.9]{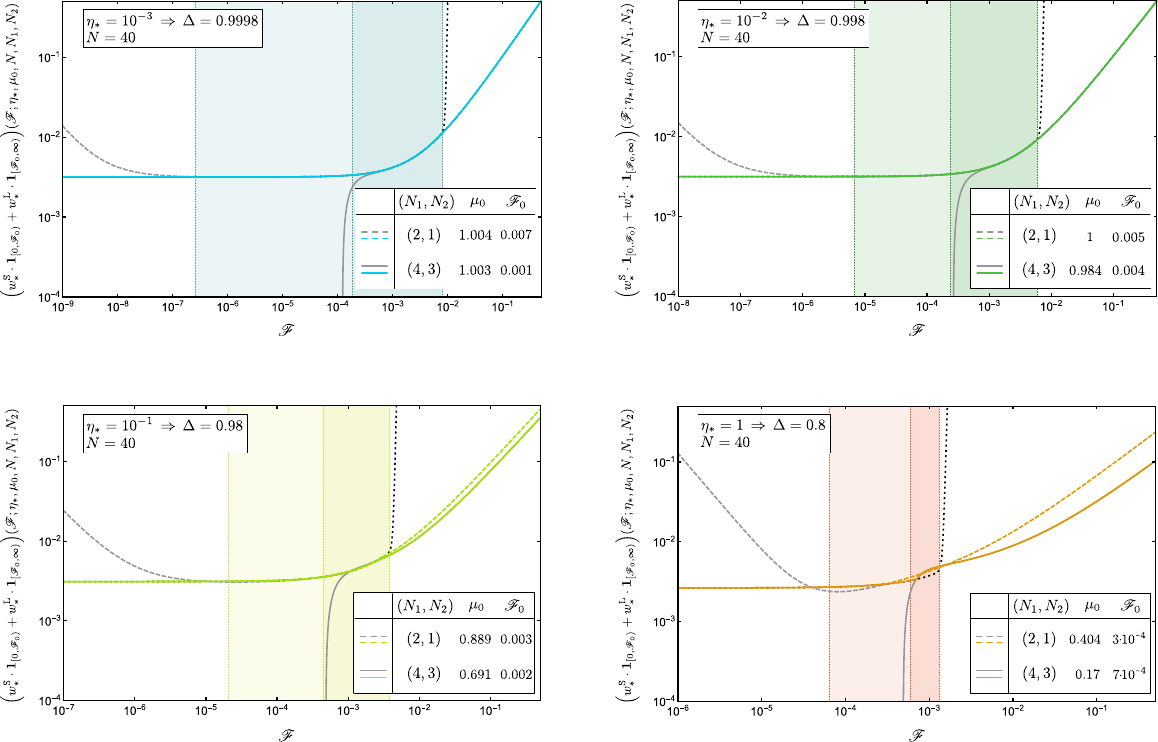}
\caption{Global continuous fixed functions and residuals of small- and
large-field expansions for various values of the anomalous dimension $\eta_{*}$
for different truncation orders. The color code for different values of
$\eta_\ast$ corresponds to that of Fig.~\ref{fig.e.a}. The large-field
asymptotics is given by a power law with exponent $\Delta =
\frac{4}{4+\eta_{*}}$, cf. upper left legends. The small-field expansion is
truncated beyond $N = 40$, whereas the large-field truncation parameters
$(N_{1},N_{2})$ are taken to be either $(2,1)$ (dashed) or $(4,3)$ (solid). In
each panel, the colored lines show combined global solutions according to
\Eqref{eq:e.p}, where the parameters $\mu_{0}$ and $\mathscr{F}_{0}$ are such
that (\ref{eq:e.o}) is fulfilled and the intersection has the least slope
difference. Numerical values are given in the boxes in the lower right corners.
Moreover, gray lines represent pure large-field solutions in both truncations,
whereby the pure small-field solution is distinguished from them by the black
dotted line in each panel. The overlap region, i.e., the interval
$[b_{\mathrm{L}},r]$ is indicated by the colored regions.}
\label{fig.e.e}
\end{center}
\end{figure*}

\textbf{(4)} As a last step, both $w_{*}^{\mathrm{S}}$ and $w_{*}^{\mathrm{L}}$
are suitably glued within the interval $[b_{\mathrm{L}},r]$. A simple procedure
is to use an intersection point of the two expansions. If more than one
intersections appear, say $\mathscr{F}_{1},\mathscr{F}_{2},\ldots$, we choose
the one for which the absolute difference $\vert
(w_{*}^{\mathrm{S}})^{\prime}(\mathscr{F}_{\ell}) -
(w_{*}^{\mathrm{L}})^{\prime}(\mathscr{F}_{\ell};\mu_{0})\vert$ is minimal to
achieve the smoothest transition possible with this direct method. Suppose we
have chosen such a point of intersection in this way, $\mathscr{F}_{\ell}$ for
some index $\ell$, then replace $\mathscr{F}_{\ell} \rightarrow \mathscr{F}_{0}$
and let $\mathscr{F}_{0} \in [b_{\mathrm{L}},r]$. With this choice, we obtain an
approximate global solution from
\begin{equation}
w_{*}(\mathscr{F}) = \Bigl( w_{*}^{\mathrm{S}}\cdot\mathbf{1}_{[0,\mathscr{F}
_{0})} +  w_{*}^{\mathrm{L}}\cdot\mathbf{1}_{[\mathscr{F}_{0},\infty)}\Bigr)
(\mathscr{F}).
\label{eq:e.p}
\end{equation}
The results for two different truncations $(N_{1},N_{2}) \in \lbrace (2,1) ,
(4,3)\rbrace$ are presented in Fig.~\ref{fig.e.e}.

Based on the quantitative results, we can first and foremost conclude that
satisfactory approximations to continuous global fixed functions can be
constructed for a variety of different $\eta_{*}$ values and truncations. With
our matching condition (\ref{eq:e.o}) we even obtain comparatively smooth
fixed functions at the fusion point $\mathscr{F}_{0}$ for sufficiently small
$\eta_\ast>0$. For the case $\eta_\ast=1$, it is evident from
Fig.~\ref{fig.e.e} that this gluing procedure leads to a visible kink at the
matching point. This may already be taken as an indication that $\eta_\ast=1$
lies beyond the range of $\eta_\ast$ values for which the present procedure
yields a valid approximation of a global fixed function. This is discussed in
more detail in the next section, where also another gluing procedure will be
presented that yields differentiable global approximations.

Let us finally comment on the convergence properties of our truncated expansion
for $N,N_{1},N_{2} \rightarrow \infty$. For this, we use the dependence of the
overlap region $[b_{\mathrm{L}},r](N,N_{1},N_{2};\eta_{*})$ on the truncation
parameters and the anomalous dimension as an indicator. For fixed $N$ and
$\eta_{*}$ we can infer from Fig. \ref{fig.e.e} that the interval obeys the
inclusion relation $[b_{\mathrm{L}},r](N,N_{1},N_{2};\eta_{*}) \supset
[b_{\mathrm{L}},r](N,N_{1}+m+1,N_{2}+m;\eta_{*})$, for $m\in\mathbb{N}$. More
generally, we expect that increasing the large-field truncation makes the
overlap region smaller for fixed $N$ and $\eta_{*}$. Increasing both $N$ and
$\eta_{*}$ in addition amplifies this effect, because it reduces the radius of convergence $r$.
Quantitatively, the overlap regions depend sensitively on $N_{1}, N_{2}$  and
$\eta_{*}$. For instance, for $\eta_{*} = 1$ at $(N_{1},N_{2}) = (4,3)$ the
overlap spans much less than one order of magnitude. For small anomalous
dimensions, the overlap is considerably larger, but also contracts sizably when
going from the $(2,1)$ truncation to the $(4,3)$ truncation. Still, the
overlap remains sufficiently large to obtain a comparatively smooth global
approximation in contrast to the $\eta_\ast=1$ case.

Of course, if the small- and/or large-field expansion is only an
asymptotic series, then the overlap region will eventually vanish for large
truncation parameters. Nevertheless, finite truncations would then still
represent quantitatively trustworthy approximations that serve to construct
global solutions. We consider the approximations constructed in the present
section to provide satisfactory evidence for the existence of a continuous
family of solutions for small values of $\eta_\ast$.

\subsection{Absence of a Movable Singularity}

The existence of a continuous family of fixed functions for small positive
$\eta_\ast$ evidenced by our preceding construction is rather unusual. While the
initial conditions, in principle, allow for a continuous solution family, the
intrinsic nonlinearity of the FFE reduces this continuous set typically to a
discrete set of solutions.

For instance for the paradigm example of scalar O($N$) models, the matching of
small- and large-field expansions imposes a condition that is only satisfied for
a finite set of solutions (typically only one solution corresponding to the
Wilson-Fisher fixed point). This can also be rephrased as follows: the proper
initial conditions at small field amplitude can also be reformulated as boundary
conditions to be imposed in the small- and large-field limit, e.g., in terms of
the potential derivative at zero field and the large-field asymptotics. While an
initial-value problem can feature continuous solution sets, a boundary-value
problem can single out discrete solutions. 

Another way to see this reduction or ``quantization'' of fixed-function
solutions goes as follows: bringing the FFE to normal form, the differential
equation, e.g., for the O($N$) model reads
\begin{equation}
    v''(\varphi) = \frac{e(v,v';\varphi)}{s(v,v';\varphi)}, \label{eq:WFnf}
\end{equation}
where $v$ denotes the potential, $\varphi$ the field amplitude, and $e$ and $s$
are functions of the potential and its first derivative. In particular, $s$
typically corresponds to the scaling term, i.e., the O($N$) analog of the
left-hand side of the FFE \eqref{eq:d.a}. For generic initial conditions imposed
at $\varphi=0$, the denominator $s$ develops a zero at some finite $\varphi$
which is called a movable singularity of the FFE. If such a movable singularity
exists, a global solution for $v(\varphi)$ can only be constructed provided that
$e$ also vanishes at this zero of $s$. This imposes another condition on the
initial values and thus leads to a quantization of solutions
\cite{Morris:1994ki,Codello:2012ec,Codello:2014yfa,Vacca:2015nta,Hellwig:2015woa}. 

For the present case, this implies that our continuous family of solutions to
the FFE for small $\eta_\ast$ persists only if the FFE \eqref{eq:d.a} does not
feature such a movable singularity. Unfortunately, this is difficult to check
directly, since we cannot bring \Eqref{eq:d.a} analytically into normal form. 

In order to collect indirect evidence, we proceed as follows: we first expand
the integrands of the threshold functions in Eqs.~\eqref{eq:threshold} in a
geometric series and perform the loop integration term by term. For instance, to
lowest nontrivial order, we obtain the differential equation
\begin{equation}
a_{0} + a_{1}\cdot(\mathscr{F}w''_{*}) + a_{2}\cdot(\mathscr{F}w''_{*})^{2} = 0,
\label{eq:nf}
\end{equation}
with coefficients
\begin{equation}
\begin{split}
&a_{0}(\mathscr{F},w'_{*};\eta_{*}) = 280 - 45\eta_{*} \\[0.5em]
&\hspace*{1cm}+ 5\left[\eta_{*} - 8 - 96\pi^{2}\left(1 - \frac{1}{\Delta(\eta_{*})}\mathscr{F}\right)\right]w'_{*}, \\[0.5em]
&a_{1}(w'_{*};\eta_{*}) = (13\eta_{*} - 124) - 2(5\eta_{*} - 54)w'_{*} \\[0.5em]
&\hspace*{1cm}+ 2(\eta_{*} - 12){w'_{*}}^{2}, \\[0.5em]
&a_{2}(w'_{*};\eta_{*}) = (5\eta_{*}-54) + 2(\eta_{*} - 12)w'_{*}. \\[0.5em]
\end{split}
\label{eq:nfcoeff}
\end{equation}
Equation \eqref{eq:nf} is a quadratic polynomial in $\mathscr{F}w''_{*}$
and can straightforwardly be brought into normal form analogously to
\Eqref{eq:WFnf}. To this order, it turns out that the resulting condition for
the absence of a movable singularity is satisfied if and only if $a_{2} \neq 0$. Conversely, if a
movable singularity is present, say at $\mathscr{F} = \mathscr{F}_{\mathrm{ms}}$, then we find
\begin{equation}
w'_{*}(\mathscr{F}_{\mathrm{ms}};\eta_{*}) = -\frac{1}{2}\frac{54-5\eta_{*}}{12-\eta_{*}}.
\label{eq:mscrit}
\end{equation} 
Since our attention is devoted to small $\eta_{*} > 0$, this expression signifies a
negative slope of the field strength potential at the movable singularity. However,
if the convergence criterion
of the geometric series expansion of the threshold functions is fulfilled, that is if
\begin{equation}
\forall \mathscr{F} \in \mathbb{R}_{0}^{+}: \quad w'_{*}(\mathscr{F}),w'_{*}
(\mathscr{F}) + \mathscr{F}w''_{*}(\mathscr{F}) \in [0,2),
\label{eq:conv}
\end{equation}
then \Eqref{eq:mscrit} cannot be true for any $\mathscr{F}$ and thus there is no
movable singularity.

In order to check this, we have to construct global solutions which are
differentiable at least twice. For this, we use an interpolation of
the small- and large-field expansions in the overlap region $[b_\text{L},r]$ by means of an affine combination:
\begin{equation}
w_{*} = g_{\mathrm{S}}w_{*}^{\mathrm{S}} + g_{\mathrm{L}}w_{*}^{\mathrm{L}} = 
w_{*}^{\mathrm{L}} + g_{\mathrm{S}}\bigl(w_{*}^{\mathrm{S}} - w_{*}^{\mathrm{L}}
\bigr),
\label{eq:weightsol}
\end{equation}
where we have used that the weight functions $g_{\mathrm{S}},g_{\mathrm{L}}$ add
up to unity;
$g_{\mathrm{S}} + g_{\mathrm{L}} = 1$.

The interpolation via $g_{\mathrm{S}}$ is constructed in such a way that the
contributions of the more reliable approximation dominates the derivatives near
the edges of the overlap region, e.g., $w_{*}^{\mathrm{S}}$ dominates near
$b_\text{L}$ and $w_{*}^{\mathrm{L}}$ dominates near $r$. This construction
therefore avoids artifacts and contaminations from the less trustworthy
approximation in the derivatives. On the level of the field-strength potential,
$g_{\mathrm{S}}$ gives full weight to the small-field expansion at $\mathscr{F}
= b_{\mathrm{L}}$, but suppresses it completely at $\mathscr{F} = r$, i.e.,
$g_{\mathrm{S}}(b_{\mathrm{L}}) = 1$ and $g_{\mathrm{S}}(r) = 0$. This
guarantees a seamless transition at the interval endpoints;
$w_{*}(b_{\mathrm{L}}) = w_{*} ^{\mathrm{S}}$ and $w_{*}(r) =
w_{*}^{\mathrm{L}}$. For concreteness we consider the following one-parameter
family of weights for any fixed overlap interval $[b_{\mathrm{L}},r]$:
\begin{equation}
g_{\mathrm{S}}(\mathscr{F};\alpha) = \frac{1}{2}\left[1 - 
\frac{\tanh\biggl(\alpha\left(\frac{\mathscr{F} - b_{\mathrm{L}}}{r-b_{\mathrm{L}}} - 
\frac{1}{2}\right)\biggr)}{\tanh\left(\frac{\alpha}{2}\right)}\right],
\label{eq:weight}
\end{equation}
with a continuous parameter $\alpha \in \mathbb{R}\setminus \lbrace 0 \rbrace$.
By varying $\alpha$ we can control the profile of the weight function between
the endpoints and regulate the transition sharpness from $w_{*}^{\mathrm{S}}$ to
$w_{*}^{\mathrm{L}}$ near the midpoint at $\mathscr{F} = \frac{1}{2}
(b_{\mathrm{L}} + r)$. Small values of $\alpha$ provide only slight deviations
from the linear weight function, whereas larger values progressively pronounce
the kink of the tanh graph. In order to avoid artifacts induced by the derivative of the weight function itself,
which could, in particular, jeopardize the convergence criteria in (\ref{eq:conv}), $\alpha$ should
not be chosen excessively large. On the other hand, for reasonably large values of $\alpha$, $g_{\mathrm{S}}$ becomes flat near the
endpoints. In this way we can neglect derivatives of $g_{\mathrm{S}}$ in the
corresponding region if $\alpha$ is not too small and confer derivatives of the
field strength potential to any order a form similar to \Eqref{eq:weightsol}.
This behavior is indeed suited for a smooth transit to the more reliable
approximations beyond the overlap interval. Hence, the discussion suggests to
find an adequate compromise between relatively flat ends and a gentle slope for
$g_{\mathrm{S}}\vert_{[b_{\mathrm{L}},r]}$. Reasonable choices for $\alpha$ are
usually of order one but can be varied by an order of magnitude.
 
Studying the convergence criteria \eqref{eq:conv}, we have verified explicitly
that our global solutions for $\eta_\ast=10^{-4},10^{-3},10^{-2},10^{-1}$
satisfy these criteria for a wide range of parameter values $\alpha$ and thus are
compatible with the absence of a movable
singularity. In addition, we have verified that this statement also holds for
the FFE to next order in the geometric-series expansion (the resulting FFE are
too extensive to be written down explicitly here). 

By contrast, the criterion for the convergence of the geometric series expansion
is violated by the solution for $\eta_\ast=1$, independently of $\alpha$. While this may
solely indicate a
failure of this expansion, we take this as an indication that the FFE may
possess a movable singularity for sufficiently large $\eta_\ast$. If so, further
solutions may still exist for discrete values of $\eta_\ast$. However together
with the fact that the directly glued solutions exhibit a kink and that the
overlap region of the expansions is rather small, we consider the present
observation as a further piece of evidence that $\eta_\ast=1$ as well as larger
values do not support a global solution to the FFE. 

We conjecture that a continuous family of global fixed-function solutions exists
for a finite interval $\eta_\ast\in (0,\eta_\text{cr})$ where the critical
anomalous dimension $\eta_\text{cr}$ lies in between $1/10$ and $1$.

\subsection{Near Critical Regime}
\label{subsec:nearcriticalregime}

Having constructed a global fixed function for nonlinear electrodynamics
in the truncated theory space, we now return to the analysis of the near
critical region for our solutions found with proper initial conditions. Of
central interest are the critical exponents of perturbations and the
classification of (ir)relevant directions.

In principle, we would have to construct eigenperturbations of the global
fixed function in order to reliably read off the eigenvalues of the stability
matrix after insertion of the global solution. In view of the complexity of the
FFE, we resort to a simpler method which we expect to give reasonable results
for the leading-order exponents: we simply use the stability matrix arising from
the small-field expansion inserting the fixed-point results for the coefficients
$u_{i,\ast}$ that we obtain from the small-field expansion using proper initial
conditions. As the latter can all be expressed as functions of $\eta_\ast$, the
stability matrix $B(\eta_{*}) = (D\beta)(\eta)\vert_{\eta = \eta_{*}}$ becomes a
function of only the anomalous dimension $\eta_{*}$.

Truncating the small-field expansion at order $N$, the stability matrix reduces
to an $N\times N$ submatrix $B^{(N)}(\eta_{*})$, the eigenvalues of which we
can determine straightforwardly in order to obtain the critical exponents
$\Theta_j^{(N)} (j \in \mathbb{N}_{0})$, cf., \Eqref{eq:stabmat}.
The latter are thus computable as
functions of $\eta_\ast$ for increasing truncation $N$.
As before, $\Theta_0^{(N)} \equiv \Theta_{0} = 4$
reflecting the canonical dimension of the vacuum energy holds independently of
the truncation. At low truncation orders, also the leading-order results for the
critical exponents can be worked out analytically. It is instructive to take a
look at the leading nontrivial exponent $\Theta_2$ associated essentially with
the coupling $u_{2,\ast}$. At order $N=2$, we have  
\begin{equation}
    \Theta_{2}^{(2)}(\eta_{*}) =
    - \frac{640  + 1360\eta_{*}- 153\eta_{*}^{2}}{20(8-\eta_{*})}.
    \label{eq:e.g}
\end{equation}
In the limit $\eta_\ast\to 0$, we rediscover $\Theta_{2}^{(2)}(0) = -4$ equaling
the canonical mass dimension of the dimensionful coupling $\bar{u}_{2}$ as it
should. For small $\eta_\ast>0$, the exponent receives small corrections. At
higher truncation order, the critical exponent can pick up an
imaginary part, so that we focus on the real part ($\Re$) in the following. For instance
at truncation order $N=3$, the expression $\Re[\Theta_{2}^{(3)}](\eta_{*})$ is
more extensive, but essentially of the form (\ref{eq:e.g}) replacing the
quadratic polynomial in the numerator by a cubic plus the square root of a
septic polynomial in $\eta_{*}$ and the denominator by $40(8-\eta_{*})^{2}$. A
similar modification applies for the transitions from $\Re[\Theta_{2}^{(3)}]$ to
$\Re[\Theta_{2}^{(4)}]$ and thereafter to $\Re[\Theta_{2}^{(5)}]$. 
Consequently, \Eqref{eq:e.g} admits a pole at $\eta_{*} = 8$, as expected: this pole must always be present for all critical exponents at each order $N$ since all couplings diverge at that parameter value.

As another check, we can make contact with our results for the critical exponents using 
improper initial conditions, as studied in Subsect.~\ref{par:fourtwo}. Therein, we found
a positive critical exponent $\Theta_{2}^{(N)}$ for all truncations studied (with
hindsight considered as an artifact of the improper initial conditions). This suggests that
$\Theta_{2}^{(N)}$ should feature a zero as a function of $\eta_\ast$. In fact, for
$\Theta_{2}^{(2)}$ we find a zero at $\eta_{*} \approx -0.448 =: \tilde{\eta}_{*}^{(2)}$,
using \Eqref{eq:e.g}. At $N=2$, larger anomalous dimensions
$\eta_{*} \in (\tilde{\eta}_{*}^{(2)} , 8)$, thus produce an irrelevant coupling
$u_{2,*} (\eta_{*})$. It becomes relevant for smaller values
$\eta_{*} < \tilde{\eta}_{*}^{(2)}$ in agreement with our findings of
Subsect.~\ref{par:fourtwo}. 

Moving to $N=3$, we can essentially observe the same behavior for
$\Re[\Theta_{2}^{(3)}]$ with the zero shifting to a larger value, $\eta_{*}
\approx -0.229 =: \tilde{\eta}_{*}^{(3)}$. In addition, we find several regions where
$\Theta_{2}^{(3)}$ switches from a real- to a complex-valued number, especially
near the pole. Hence, $\Re[\Theta_{2}^{(3)}]$ has discontinuities at these
points. Whenever $\Theta_{2}^{(3)}(\eta_{*}) \in \mathbb{C}$
in these regions, then, by the complex conjugate root theorem, also its complex
conjugate $\overline{\Theta_{2}^{(3)}(\eta_{*})}$ must be an eigenvalue. Now, because $\Theta_{0}^{(N)} = 4$ is true for all $N
\in \mathbb{N}_{0}$, we must have $\overline{\Theta_{2}^{(3)}} =
\Theta_{3}^{(3)}$, i.e., the critical exponents belonging to $u_{2,*}$
and $u_{3,*}$ must combine to complex conjugate pairs. The situation only
marginally changes when we increase the truncation to $N=4$ and $N=5$. The
switching behavior of $\Theta_{2}^{(4)}$ and $\Theta_{2}^{(5)}$ between
$\mathbb{R}$ and $\mathbb{C}$ is unpredictably chaotic. On the other hand, the
zeroes $\tilde{\eta}_{*}^{(4)} \approx -0.13$ and $\tilde{\eta}_{*}^{(5)} \approx -0.079$
where $u_{2,*}$ becomes relevant move closer to $0$. For reasons of continuity,
we do not expect that this sequence of zeroes,
$(\tilde{\eta}_{*}^{(N)})_{N\in\mathbb{N}}$, crosses zero, where $\Theta_{2}^{(N)}(0)=-4$
must hold to all orders. The continuity assumption hence implies that
$\tilde{\eta}_{*}^{(N)} < 0$ for all $N \in \mathbb{N}$. In this scenario, $u_{2,*}$
would be an irrelevant coupling for all positive
$\eta_{*}$, at least sufficiently below the pole at $\eta_\ast=8$. Whether this
continuity scenario applies to all orders remains an open question.

For even larger $N \geq 6$, no elementary closed-form solutions to the
characteristic polynomial of $B^{(N)}(\eta_{*})$ for arbitrary $\eta_{*}$ exist
according to the Abel-Ruffini theorem. Therefore, we continue with specific
$\eta_{*}$ values as done in Figs.~\ref{fig.e.a} and \ref{fig.e.b}, and
discuss some properties of the full spectra $\mathrm{eig}(B^{(N)}(\eta_{*}))$ up to $N
= 26$.

\begin{figure}
\begin{center}
\includegraphics[scale=0.55]{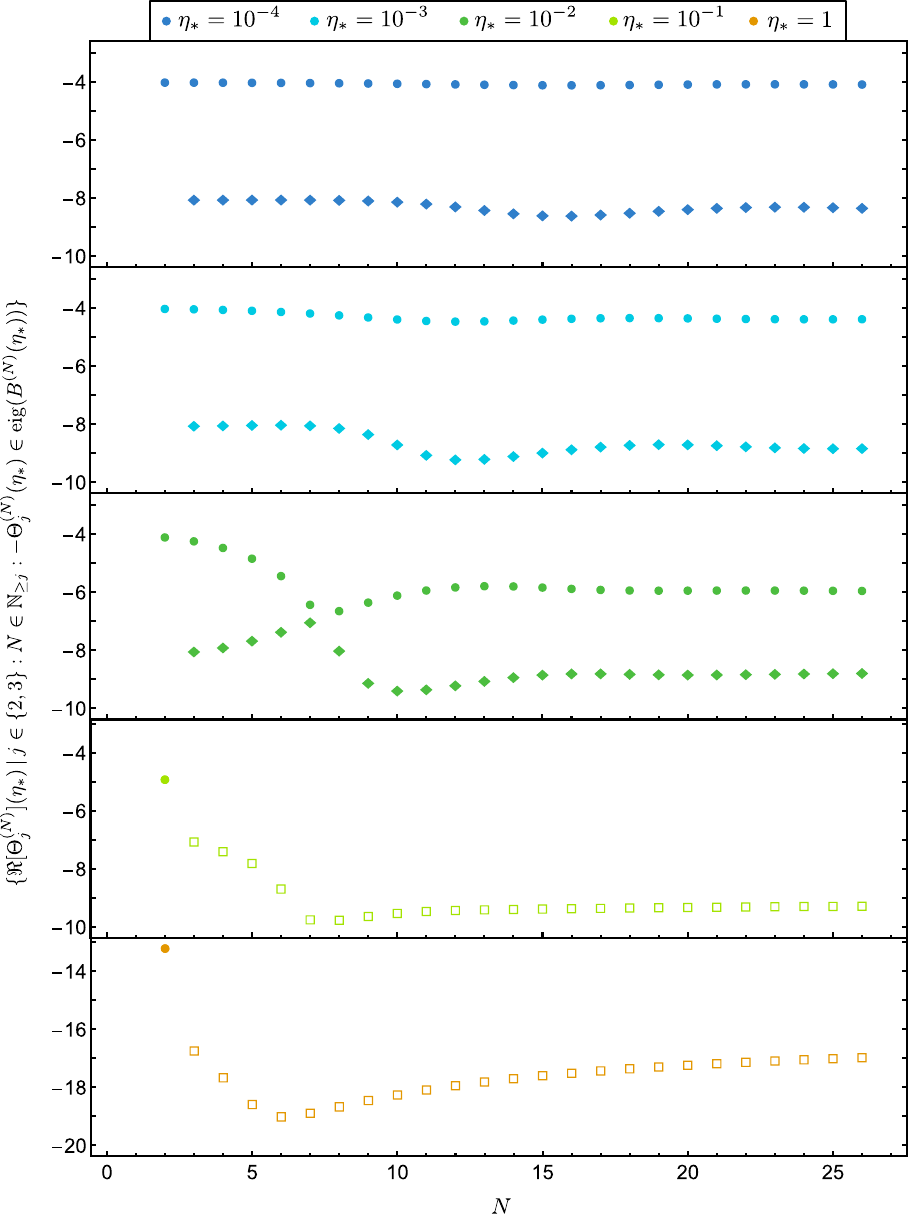}
\caption{Real parts of the leading nontrivial critical exponents  $\Re[\Theta_{2}^{(N)}]$ (filled circles) and $\Re[\Theta_{3}^{(N)}]$ (filled diamonds) corresponding to $u_{2,*}$ and $u_{3,*}$, respectively, as a function of the truncation order $N$ up to $N=26$ for a logarithmic selection of anomalous dimension values $\eta_{*}$ (color code on top of diagram). Complex pairs of critical exponents are marked by open squares. Note the shifted scale of the vertical axis in the lowest plot.}
\label{fig.e.c}
\end{center}
\end{figure}

Let us start with the first nontrivial critical exponents $\Theta_{2}^{(N)}$ and $\Theta_{3}^{(N)}$. Their $\eta_{*}$ dependence and evolution for an increasing dimension of theory space $N$ is presented in Fig.~\ref{fig.e.c}.
For small $\eta_{*}$, both $u_{2,*}$ and $u_{3,*}$ describe irrelevant
couplings and are numerically close to the canonical mass dimension of their
respective dimensionful versions with minimal variation
throughout various $N$. As $\eta_{*}$ gets larger, both of the real parts approach each other, which is particularly noticeable for small $N$, before they
eventually combine to a complex conjugate pair. Most importantly, in either case
both real parts seem to converge in the large-$N$ limit and thus indicate a
well-defined critical structure for the fixed functions constructed from a small-field expansion up to the third-order operator $\mathscr{F}^{3}$.

Unfortunately, the small-field expansion technique used here for an estimate of the leading critical exponents fails for even higher exponents $\Theta_{4}^{(N)},\Theta_{5}^{(N)}, \ldots$. Of course, for any finite $N$, estimates for the exponents are computable, but we do not observe any sign of convergence even for large values of $N$. A determination of the spectrum to higher orders thus appears to require a global study of the perturbations. E.g., using an ansatz $w_{k}(\mathscr{F}) = w_{*}(\mathscr{F}) + \mathrm{e}^{-\Theta t} \delta w(\mathscr{F})$, a linearization of the flow to leading order in $\delta w(\mathscr{F})$ can give access to the spectrum of the resulting differential operator and thus to all eigenvalues $\Theta$. However, this goes beyond the analytical methods concentrated on in the present work. 

In summary: the accessible part of the leading critical exponents covers the trivial exponent (leaving aside the superscript $(N)$) $\Theta_0=4$ for the vacuum energy, as well as the two leading nontrivial ones $\Theta_2$ and $\Theta_3$. The latter are close to their canonical values $-4$ and $-8$, respectively, for small $\eta_\ast>0$, and are shifted to even more negative values for increasing $\eta_\ast$. Assuming that this pattern holds also for the subleading critical exponents $\Theta_{j\geq 4}$, we find indications that all nontrivial exponents are negative. This implies that all nontrivial perturbations of the fixed point are RG irrelevant, i.e. the fixed function is fully attractive in the long-range limit. The physical implications are discussed below.

\section{Exploring the Full Nonlinear System at Leading-Derivative Order}
\label{sec:six}
The flow of the full nonlinear system at leading-derivative order is described
by \Eqref{eq:c.k}; the corresponding fixed points satisfy the FFE
(\ref{eq:c.m}). In the preceding sections we have specialized to the reduced
system characterized by the two approximations of self duality \textbf{(A1)} and
the exclusion of $\mathscr{G}^{2}$ dependencies \textbf{(A2)}. Let us now check
the validity of these approximations, by exploring the corrections arising from
the inclusion of $\mathscr{G}^{2}$ contributions to the flow. For this, we now
drop the approximation \textbf{(A2)}, but keep \textbf{(A1)} in order to exploit
the simplicity arising from self-duality for the operator traces. This suffices
to include the contributions from the additional operators to the flow and
monitor their quantitative relevance.

For a convenient quantitative comparison, we go back to the improper initial conditions and employ the small-field expansion. While the resulting fixed-point candidates presumably are artifacts of the truncation, they allow us to quantify the influence of the  $\mathscr{G}^{2}$-dependent terms. More specifically, we study the following series of increasing truncations:
\begin{equation}
    \begin{split}
    \mathrm{T1}: \quad w_{k}^{(1)}(\mathscr{F},\mathscr{G}^{2}) &= c_{k} + \mathscr{F} +
    \frac{1}{2}m_{1,k}\mathscr{F}^{2}\\[0.5em]&\hspace*{2.05cm} + \frac{1}{2}m_{2,k}
    \mathscr{G}^{2}, \\[0.5em]
    \mathrm{T2}: \quad w_{k}^{(2)}(\mathscr{F},\mathscr{G}^{2}) &= w_{k}^{(1)}
    (\mathscr{F},\mathscr{G}^{2}) \\[0.5em] &+ \frac{1}{2}\sigma_{1,k}\mathscr{F}^{3} + \frac{1}{2}
    \sigma_{2,k}\mathscr{F}\mathscr{G}^{2}, \\[0.5em]
    \mathrm{T3}: \quad w_{k}^{(3)}(\mathscr{F},\mathscr{G}^{2}) &= w_{k}^{(2)}(\mathscr{F},\mathscr{G}^{2}) \\[0.5em]&\hspace*{-1.9cm} + \frac{1}{3}
    \lambda_{1,k}\mathscr{F}^{4} + \frac{1}{3}\lambda_{2,k}\mathscr{F}^{2}
    \mathscr{G}^{2} + \frac{1}{3}\lambda_{3,k}\mathscr{G}^{4}.
    \end{split}
    \label{eq:f.a}
\end{equation}
In T1 we account for the effects of flow contributions proportional to
$\dot{w}_{k}$, where in T2 we also include nontrivial mixed-derivative inputs
proportional to $\dot{w}_{k}^{\prime}$. Finally, T3 considers nonvanishing
derivatives of $w_{k}$ to all occurring orders in \Eqref{eq:c.k}.

We emphasize that the self-duality approximation \textbf{(A1)} is used only after we have performed all functional derivatives to obtain the Hessian of the action. For instance on the operator level at quartic order in the field strength, we have both terms  $\frac{1}{2}m_{1,k}\mathscr{F}^{2} + \frac{1}{2}m_{2,k}\mathscr{G}^{2}$. The evaluation of the final traces using  \textbf{(A1)} then corresponds to a projection in coupling space,  $m_{1,k},m_{2,k} \mapsto m_{k}$, since $\mathscr{F}^{2}$ and $\mathscr{G}^{2}$ are identified. 

Using the abbreviation
$\rho_{q,k}(\eta_{k}) := (2q-\eta_{k})$ for $q \in \mathbb{Q}$,
the beta functions for truncation T3 read:
\begin{equation}
\begin{split}
\partial_{t}c_{k} &\equiv \beta_{c} =  -4c_{k} + \frac{\rho_{3,k}}{48\pi^{2}},
\\[1.5em]
\partial_{t}m_{k} &\equiv \beta_{m} = -2\rho_{-1,k}m_{k} \\[0.5em]
&\hspace*{1.6cm}+ \frac{1}{640\pi^{2}}\Bigl(11\rho_{5,k}m_{k}^{2} 
- 20\rho_{4,k}\sigma_{k}\Bigr), \\[1.5em]
\partial_{t}\sigma_{k} &\equiv \beta_{\sigma} = -3\rho_{-\frac{4}{3},k}\sigma_{k} 
- \frac{1}{960\pi^{2}}\Bigl(29\rho_{6,k}m_{k}^{3} \\[0.5em]&\hspace*{2.4cm}
 - 84\rho_{5,k}m_{k}\sigma_{k} + 45\rho_{4,k}\lambda_{k}\Bigr),\\[1.5em]
\partial_{t}\lambda_{k} &\equiv \beta_{\lambda} = -4\rho_{-\frac{3}{2},k}
\lambda_{k} + \frac{1}{6720\pi^{2}}\Bigl(415\rho_{7,k}m_{k}^{4}
\\[0.5em]&\hspace*{-0.5cm}  - 1596\rho_{6,k}m_{k}^{2}\sigma_{k}
+ 756\rho_{5,k}\sigma_{k}^{2} + 966\rho_{5,k}m_{k}\lambda_{k}\Bigr).
\end{split}
\label{eq:f.b}
\end{equation}
The corresponding beta functions for truncations T2 and T1 can be
inferred from (\ref{eq:f.b}) by setting $\lambda_{k} = 0$ for T2 and
additionally $\sigma_{k} = 0$ for T1. Furthermore, the scale-dependent
anomalous dimension is given by
\begin{equation}
\eta_{k}(m_{k}) = \frac{10\left(\frac{m_{k}}{2}\right)}{48\pi^{2} + \frac{5}{4}
\left(\frac{m_{k}}{2}\right)}.
\label{eq:f.c}
\end{equation}

At a fixed point, the anomalous dimension is only marginally modified
compared to \Eqref{eq:SMEtower} upon identifying $\frac{m_{*}}{2}$ with $u_{2,*}$.
This identification indeed is consistent: the approximations \textbf{(A2)} and
\textbf{(A1)} (performed in this order) reduce the quadratic term of each truncation T1,
T2, and T3 to $\frac{m_{*}}{2}\mathscr{F}^{2}$ at a fixed point. In the same way,
$u_{3,*}$ can be identified with $\frac{\sigma_{*}}{2}$ and $u_{4,*}$ with
$\frac{\lambda_{*}}{3}$. Interestingly, $\eta_{*}(m_{*})$ is obtained exactly from the
expression in (\ref{eq:SMEtower}) by replacing $u_{2,*}$ with $1.2u_{2,*}$.

In Table~\ref{tab:c.b}, we list all fixed-point values appearing in each
truncation. Table~\ref{tab:c.c} displays the corresponding anomalous dimension and the critical exponents. From this data, we deduce the following results:

\begin{table}[t]
\centering
\renewcommand{\arraystretch}{2}
\begin{tabular}{|c|c|c|c|c|}
\hline
Truncation & $c_{*}$ & $\frac{m_{*}}{2}$ & $\frac{\sigma_{*}}{2}$ & $\frac{\lambda_{*}}{3}$ \\
\hline\hline
T1 & \begin{tabular}{c}
$0.0032$ \\
\hline $0.0038$
\end{tabular} & \begin{tabular}{c}
$0$ \\ \hline $-46.108$
\end{tabular} & $/$ & $/$ \\
\hline
T2 & \begin{tabular}{c}
$0.0032$ \\ \hline $0.0035$ \\ \hline $0.0041$
\end{tabular} & \begin{tabular}{c}
$0$ \\ \hline $-25.631$ \\ \hline $-66.245$
\end{tabular} & \begin{tabular}{c}
$0$ \\ \hline $-1787.8$ \\ \hline $4504.7$
\end{tabular} & $/$  \\
\hline
T3 & \begin{tabular}{c}
$0.0032$ \\ \hline $0.0033$ \\ \hline $0.0038$ \\ \hline $0.0042$
\end{tabular} & \begin{tabular}{c}
$0$ \\ \hline $-14.657$ \\ \hline $-48.181$ \\ \hline $-76.839$
\end{tabular} & \begin{tabular}{c}
$0$ \\ \hline $-1573.9$ \\ \hline $338.58$ \\ \hline $7955.8$
\end{tabular} & \begin{tabular}{c}
$0$ \\ \hline $-107521$ \\ \hline $249995$ \\ \hline $-523227$
\end{tabular} \\
\hline
\end{tabular}
\caption{Fixed-point coordinates for increasing truncations T1, T2, and T3 of theory space of full self-dual nonlinear electrodynamics at leading derivative order using conventional improper initial conditions.}
\label{tab:c.b}
\end{table}
\begin{table}[t]
\centering
\renewcommand{\arraystretch}{2}
\begin{tabular}{|c|c|c|c|c|c|}
\hline
Truncation & $\eta_{*}$ & $\Theta_{c}$ & $\Theta_{m}$ & $\Theta_{\sigma}$ & $\Theta_{\lambda}$ \\
\hline\hline
T1 & \begin{tabular}{c}
$0$ \\ \hline $-1.108$
\end{tabular} & \begin{tabular}{c}
$4$ \\
\hline $4$
\end{tabular} & \begin{tabular}{c}
$-4$ \\ \hline $4.5096$
\end{tabular} & $/$ & $/$ \\
\hline
T2 & \begin{tabular}{c}
$0$ \\ \hline $-0.5803$ \\ \hline $-0.6556$
\end{tabular} & \begin{tabular}{c}
$4$ \\ \hline $4$ \\ \hline $4$
\end{tabular} & \begin{tabular}{c}
$-4$ \\ \hline $4.2625$ \\ \hline $4.8249$
\end{tabular} & \begin{tabular}{c}
$-8$ \\ \hline $-5.2259$ \\ \hline $14.919$
\end{tabular} & $/$  \\
\hline
T3 & \begin{tabular}{c}
$0$ \\ \hline $-0.3218$ \\ \hline $-1.1652$ \\ \hline $-2.0344$
\end{tabular} & \begin{tabular}{c}
$4$ \\ \hline $4$ \\ \hline $4$ \\ \hline $4$
\end{tabular} & \begin{tabular}{c}
$-4$ \\ \hline $4.1465$ \\ \hline $4.5389$ \\ \hline $5.0627$
\end{tabular} & \begin{tabular}{c}
$-8$ \\ \hline $-5.2411$ \\ \hline $-5.4326$ \\ \hline $10.636$
\end{tabular} & \begin{tabular}{c}
$-12$ \\ \hline $-11.065$ \\ \hline $19.2$ \\ \hline $33.336$
\end{tabular} \\
\hline
\end{tabular}
\caption{Anomalous dimension and critical exponents at the fixed points of
Table~\ref{tab:c.b}. The order of the associated fixed points, read
from top to bottom, is the same as given in Table~\ref{tab:c.b}.}
\label{tab:c.c}
\end{table}

\textbf{(1)} As a zeroth-order check, also the full system shows a Gaussian
fixed point which is characterized by $\eta_{*} = 0$ and where all couplings
except for the constant $c_{*}$ vanish identically. The latter takes the same
value as in the \textbf{(A2)} and \textbf{(A1)} reduced system. Therefore, the
Gaussian fixed function is still and, in fact, exactly, given by
$w_{\mathrm{GFP}}(\mathscr{F}, \mathscr{G}^{2}) = w_{\mathrm{GFP}}(\mathscr{F})
= \frac{1}{32\pi^{2}} + \mathscr{F}$. The critical exponents agree with the
canonical mass dimensions of corresponding dimensionful couplings.

\textbf{(2)} In addition to the Gaussian fixed point, we find further
non-Gaussian fixed points, the number of which increases from truncation T$N$ to
T$(N+1)$ exactly in the same way as in our previous analysis of Sect.
\ref{sec:four} for the reduced FFE. A direct comparison between the
NGFP\textbf{1} branch of Fig.~\ref{fig.d.a} and the $\frac{m_{*}}{2}$ column of
Table~\ref{tab:c.b} reveals an initial relative numerical deviation of about
$8$\% while qualitatively showing the same falloff behavior towards the
Gaussian fixed point as we increase the truncation order. The same can be seen
for the NGFP\textbf{2} branch and is expected to continue for higher-order
non-Gaussian fixed-point classes. Since it is still possible to express each
coupling $\sigma_{*}, \lambda_{*}, \ldots$ as a function of $m_{*}$, small
deviations between $\frac{m_{*}}{2}$ and $u_{2,*}$ imply likewise small
deviations between $\frac{\sigma_{*}}{2}$ and $u_{3,*}$, $
\frac{\lambda_{*}}{3}$ and $u_{4,*}$, etc. The same argument applies to the
anomalous dimension.

\textbf{(3)} For the critical exponents, we essentially obtain the same picture
as in Subsect.~\ref{par:fourtwo}. According to Table~\ref{tab:c.c}, the
fixed-point class NGFP\textbf{n} exhibits $n+1$ relevant directions.
Moreover, $\Theta_{m}$ shows only a minor numerical discrepancy with our
previous findings, cf. Fig.~\ref{fig.d.d}.

In summary, we found that the additional $\mathscr{G}^{2}$-dependent terms in
the full nonlinear theory to leading-derivative order contribute only
quantitatively marginal effects in comparison to the purely
$\mathscr{F}$-dependent description of the fixed-point action and the near
critical regime. This conclusion holds at least for the small-field expansion
using the conventional improper initial conditions. Here, we do not observe any
relevant changes in the system's overall behavior and conclude that the
assumption \textbf{(A2)} serves a legitimate and efficient approximation in
addition to self-duality. It demonstrates the self-consistency of our
geometrical line of argument discussed in Subsect.~\ref{par:threefour}.

\section{Interpretation Scenarios}
\label{sec:seven}

Let us assume that our nonperturbative results observed for nonlinear
electrodynamics also hold beyond the leading-derivative order. Of course, while
corrections from higher derivative order are guaranteed to feed back into the
lower orders, their contributions to the near critical regime can be expected to
remain power-counting suppressed by their higher canonical dimension. This
statement is exact at the Gaussian fixed point; and since our solutions for
proper initial conditions feature a small anomalous dimension, we expect
power-counting arguments to be reliable also in this immediate vicinity of the
Gaussian fixed point. 

In the following, we discuss several interpretation scenarios. All scenarios are
based on the global fixed-point solutions which we found for proper initial
conditions, but differ due to additional assumptions or the inclusion of further
degrees of freedom.

\subsection{UV-Complete Nonlinear Electrodynamics}

The existence of further non-Gaussian fixed points in theory space is a
prerequisite for the asymptotic-safety scenario. The global fixed functions
which we constructed with proper initial conditions with a positive $\eta_\ast$
can then be viewed as scaling solutions of a continuous set of UV fixed points.
Each fixed point defines a universality class of UV-complete theories of
nonlinear electrodynamics. The long-range behavior in each universality class is
then governed by the RG relevant directions of each fixed point.

Interestingly, the only relevant direction is given by the vacuum energy
according to our results of Subsect.~\ref{subsec:nearcriticalregime}, all other
directions for which we have reliable data are RG irrelevant. This implies that
the nontrivial part of the fixed functional is also IR attractive. For the theory initiated on an RG trajectory emanating from the fixed point in the UV, it remains on the quantum-scale-invariant
solution over all scales; in other words, it never leaves the fixed point. The scaling solution, parametrizing the effective action in the form of a nontrivial fixed function, thus governs also the nonlinear interactions of the long-range physics.
Apart from the trivial vacuum energy to be fixed by a renormalization condition
(and ignored in the following discussion), the full quantum effective action in
the low-energy limit is thus given by $\Gamma_{k_0}[A]= \int_{\mathbb{R}^{4}}
w_{\ast}(\mathscr{F},\mathscr{G}^{2})\mathrm{d}^{4}s$, where $s$ is a dimensionless
integration variable and $k_0$ means some fiducial low-energy
scale that serves as a measurement scale for all dimensionful quantities such as
the field amplitudes, and $\mathscr{F},\mathscr{G}^{2}$ are understood to be
measured in units of this scale. Such a theory then has no free parameter and is maximally predictive. In the present scenario, the value of $\eta_\ast$ does not play the role of a parameter of the theory, but rather characterizes different theories each forming a universality class labeled by $\eta_\ast$.

In this scenario, it remains a question as to whether this low-energy theory is
a genuinely nontrivial theory. For instance, in the reduced version where we
have studied only the dependence on the invariant $\mathscr{F}$, the effective
Lagrangian $\mathcal{L}(\mathscr{F}) = w_{\ast}(\mathscr{F},0)$ is a positive and monotonic
function in the Euclidean. This suggests that we could perform a nonlinear field
transformation of the classical fields, i.e., the expectation values of the
quantum fields, $A\to\hat{A}$, such that $\mathcal{L}(\mathscr{F}) =
w_{\ast}(\mathscr{F},0)\equiv \hat{\mathscr{F}}$ assumes the form of the
noninteracting Maxwell Lagrangian for the transformed gauge field $\hat{A}$.
However, since our results of Sect.~\ref{sec:six} suggest that the scaling
solution also depends nontrivially on the invariant $\mathscr{G}$, a
transformation $ w_{\ast}(\mathscr{F},\mathscr{G}^2) \to \hat{\mathscr{F}}$ does
most likely not exist. In this case, the scaling solution does represent a nontrivial interacting theory on macroscopic scales.

It is instructive to study the effective action also in Minkowski space. Using
$k_0$ as an IR reference scale, the effective Lagrangian expressed in terms of
the dimensionless Minkowski-valued invariants reads to lowest order:
\begin{equation}
    \frac{\bar{\mathcal{L}}(\bar{\mathcal{F}},\bar{\mathcal{G}}^{2})}{k_0^4} = - \mathcal{F} 
    - \frac{1}{2} m_{1} \mathcal{F}^2 
    + \frac{1}{2} m_{2} \mathcal{G}^2. 
\end{equation}
This form of the Lagrangian is well known from the weak-field analysis of the
Heisenberg-Euler action \cite{Heisenberg:1936nmg}. The leading-order nonlinear
coefficients $m_{1,2}$ can be related to the properties of light propagation in
an external field,
cf.~\cite{Bialynicka-Birula:1970nlh,Adler:1971wn,Dittrich:2000zu}. Now,
causality can be argued to impose constraints on the values of these
coefficients \cite{Shore:1995fz,Shore:2007um}; more precisely, requiring that
the phase velocities of low-energy photons do not exceed the vacuum speed of
light, implies \cite{Shore:1995fz,Dittrich:2000zu,Shore:2007um}
\begin{equation}
    m_{2}-m_{1}\geq 0. \label{eq:causality}
\end{equation}
While the more relevant quantity for causality actually is the front velocity
which is given by the high-frequency limit of the phase velocity, it has been
argued that the front velocity is always bound from below by the low-frequency
phase velocity for a nonamplifying ground state
\cite{Drummond:1979pp,Barton:1992pq,Shore:2007um}. Therefore, our resulting
effective action needs to satisfy the causality constraint \Eqref{eq:causality}.
For the present truncation, we find $m_{2}-m_{1}=0$ as a result of the self-dual
approximation rather than of a full calculation. Whether or not the fixed-point
action does satisfy all necessary causality constraints hence requires further
investigation going beyond the self-dual techniques used in the present work.  

In summary, we conclude that our fixed functions constitute a UV-complete
version of nonlinear electrodynamics which is essentially parameter free, as the
long-range physics is also governed by the scaling solution. We emphasize that
this scenario does not solve the triviality/Landau-pole problem of QED, since
the latter arises from interactions with matter which are not included in the
present scenario. 

\subsection{Low-Energy QED Effective Action}

Within the context of QED, the inclusion of electron matter degrees of freedom
modifies the picture for pure nonlinear electrodynamics in several ways. First,
electron fluctuations induce an anomalous dimension for the gauge field. In QED, we
have at one-loop order $\eta=\frac{2}{3\pi} \alpha$. For $\alpha \approx 1/137$, this
yields $\eta\approx 0.00155$. This value is well within the regime where we have been
able to construct global fixed functions.

A second modification arises from the fact that the electronic fluctuations, of course,
also contribute to the flow of the effective action. On the level of the flow equation,
this contribution serves as an inhomogeneous source term, depending on the gauge
field, but not on the field-strength potential $w_k$. For instance, integrating only this
source contribution leads to the one-loop Heisenberg-Euler effective action
\cite{Heisenberg:1936nmg}. From a perturbative viewpoint, the contributions arising
from integrating the flow induced by the terms depending nonlinearly on $w_k$ and
its derivatives correspond to resumming higher-loop contributions. In the perturbative
domain, these terms are subleading compared to the one-loop terms at least in the
small-field domain. 

The situation is less clear at large field amplitudes where the size of the amplitude can
make up for a small-coupling value. If $\eta\approx\text{const.}$ for a sizable number of
scales, the IR attractive nature of our fixed-point solution can win out over the matter
induced direct terms. In this case, we expect the Minkowskian-valued effective
Lagrangian at some low-energy scale $k_0$ to assume the asymptotic strong-field
form, cf. \Eqref{eq:e.j},
\begin{equation}
 \frac{\bar{\mathcal{L}}(\bar{\mathcal{F}},\ldots)}{k_0^4}\sim - \mathcal{F}^{\Delta(\eta)} \quad (\bar{\mathcal{F}} \rightarrow \infty), \quad \Delta(\eta) = \frac{4}{4+\eta}.    
 \label{eq:QEDlargefield}
\end{equation} 
Using $\eta=\frac{2}{3\pi} \alpha$ and expanding \Eqref{eq:QEDlargefield} in
powers of $\alpha$ for the example of a magnetic background $\mathsf{B}$, we
obtain to order $\alpha$ at large fields:
\begin{equation}
    \frac{\bar{\mathcal{L}}(B)}{k_0^4} \bigg|_{\alpha} \sim  \frac{\alpha}{6\pi} \mathsf{B}^2 \ln\bigl( \mathsf{B}\bigr).
\end{equation}
Identifying $k_0$ with the electron mass scale, $k_0=m_e$, this result
corresponds precisely to the strong magnetic field limit of the one-loop
Heisenberg-Euler Lagrangian. Incidentally, if we had used the correct two-loop
anomalous dimension, we would have found the correct two-loop strong-field limit
in line with an argument relating the strong-field limit with the trace anomaly
of the energy-momentum tensor \cite{Pagels:1978dd,Dunne:2002ta}.

While we consider this observation as reassuring for our result for the
strong-field limit, note that the argument based on the trace anomaly relies on
identifying the RG scale with the strong-field scale. This identification is
also used for leading-log resummations of the perturbative strong-field series
\cite{Ritus:1977iu,Dittrich:1985yb}. In complete analogy to the leading-log
resummation for the running coupling, such resummations lead to a Landau-pole
singularity in the effective action at exponentially large field strength
$\bar{\mathsf{B}} \sim m_{e}^2 e^{3\pi/\alpha}$. 

By contrast, the resummation implicitly performed by the functional RG flow
equation does not rely on scale identification. Thus, there is no
reason for the Landau pole of the running coupling at high momenta to be
translated into a similar singularity for large field strength. In fact, the
asymptotic form suggested by \Eqref{eq:QEDlargefield} as well as our full global
solutions are free of any singularity. Therefore, we conjecture that
\Eqref{eq:QEDlargefield} describes the strong magnetic field limit of the 1PI
effective action of QED. We emphasize that this is not in contradiction with
strong-field results for the Heisenberg-Euler action based on the Schwinger
functional dominated by one-particle reducible (1PR) diagrams
\cite{Karbstein:2019wmj,Karbstein:2021gdi}. 

More precisely, we consider \Eqref{eq:QEDlargefield} to hold as long as the
magnetic field is the dominating scale with all other energy scales (test
particles, photons, etc.) being much smaller and in the perturbative domain.
Also, \Eqref{eq:QEDlargefield} does not hold for the equivalent electric case
which is dominated by Schwinger pair production and an energy transfer to
particle degrees of freedom. We emphasize that the singularity-free strong-field
limit \Eqref{eq:QEDlargefield} does also not solve the Landau pole problem nor
render QED UV complete, since the high-energy behavior remains dominated by
$\eta$ growing large (with a Landau-pole singularity as a perturbative artifact)
and thus a UV limit being precluded by a causally disconnected chirally broken
phase \cite{Gockeler:1997dn,Gies:2004hy}.

As a last comment, it is tempting to speculate if our fixed functional may have
any relevance for the deep IR limit of QED far below the electron mass
threshold, since the matter induced photon self-interactions render the
effective theory an interacting one. However, at the same time, the leading
matter dominated contributions to the anomalous dimension decouple below the
electron mass threshold, $\eta|_{k\ll m_e}\to 0$, such that the mechanisms
leading to our nontrivial solutions disappear towards the deep IR. Of course, a
more precise answer requires a full (numerical) solution of the RG flow for
$w_k$ including the matter source terms and the electron decoupling. 

\subsection{High-Energy Hypercharge Sector of the Standard Model}

In the full standard model, our results may find application for the hypercharge sector
above the electroweak scale. Here, the anomalous dimension  at one-loop order is
given by $\eta_Y= \frac{41}{3} \frac{g_Y^2}{(4\pi)^2} $, cf.
\cite{Bohm:2001yx}. At the electroweak scale, we have $g_Y\approx 0.36$ for a NNLO
fixing at the top mass scale \cite{Buttazzo:2013uya} increasing mildly to
$g_Y\approx 0.48$ at about the Planck scale. This corresponds to values of the
anomalous dimension of about $\eta_Y\approx 0.01 \dots 0.02$ and thus well in the
range of values for which we find global solutions with a strong-field asymptotic limit
given by \Eqref{eq:QEDlargefield} in the magnetic field direction.

Our conclusion for this application is therefore similar to that for QED discussed in the
preceding subsection: provided the global fixed function is sufficiently attractive under
the RG flow, the 1PI effective action exists globally for any value of the hypercharge
magnetic field.

\section{Conclusions}
\label{sec:conc}

We have investigated the renormalization flow in the theory space of abelian
quantum gauge fields, i.e., nonlinear electrodynamics, using the nonperturbative
functional RG. For this, we have concentrated on the RG flow of the 1PI
effective action to leading order in a derivative expansion. The resulting flow
equation \eqref{eq:c.k} for the effective Euclidean Lagrangian or field-strength
potential represents a main result of our work and generalizes a previous result
for the magnetic theory space \cite{Laporte:2022ziz} to the full space of
nonderivative invariants $\mathscr{F}$ and $\mathscr{G}^{2}$. 

In order to explore the phase diagram in this theory space on nonlinear
electrodynamics, we focus on the possible existence of fixed points and
construction of corresponding fixed functionals. Most of our explicit results
refer to an approximation scheme relying on simplifications for a self-dual
choice of the electromagnetic field; in addition, we provide evidence that our
results are only mildly modified if these approximations are dropped. For the
construction of fixed-point actions, we have used two approaches: The first one
corresponds to a conventionally used small-field expansion. This procedure
corresponds to an implicit use of improper initial conditions for the
fixed-point equation yielding a large number of artifact fixed points. We
observe a pattern different from Wilson-Fisher-like systems but similar to that
discovered for shift-symmetric theories \cite{Laporte:2022ziz,deBrito:2023myf}.
Following \cite{deBrito:2023myf}, we arrive at the conclusion that none of the
nontrivial fixed points using improper initial conditions approximates a valid
fixed point in full theory space.  

In the second approach, we have used proper initial conditions and studied
small- and large-field expansions. Emphasizing the importance of the criterion of
global existence, i.e., the absence of singularities of the effective action in
amplitude space, we have been able to construct full fixed functionals. In fact,
our construction provides evidence for the existence of a continuous family of
fixed points as a function of the anomalous dimension $\eta_\ast>0$ of the field
amplitude. This quantity parametrizes naturally the initial condition of the
fixed-function equation at small field amplitude as well as governs the
large-field asymptotics. We have also checked whether a quantization of the
fixed-function solutions is induced by the presence of a movable singularity in
the fixed-function equation. Our results are compatible with the absence of a
movable singularity thus facilitating the presence of a continuous solution
family. Still our results suggest the existence of a critical value of
$\eta_\ast$ above which no valid solution exists. 

Finally, we have explored the critical region of the fixed-point solutions. Our
results are compatible with this continuum of fixed points being fully
IR-attractive apart from the trivial vacuum energy. However, a more definite
conclusion requires also a global analysis of the perturbations around the fixed
point beyond our analysis so far. 

We have discussed several scenarios for which our results could be relevant.
Imminently, these non-Gaussian fixed points can serve to define an interacting
theory of nonlinear electrodynamics without matter with the fixed-point action
being equivalent to the long-range action as the fixed point is fully IR
attractive. We also offer several applications to scenarios that include matter
as the source of a nontrivial anomalous dimension. In this cases, our
fixed-point action has the potential to dominate the magnetic strong-field limit
of the effective action. We consider this an attractive feature as it removes
the puzzle of a Landau-pole type singularity in the leading-log resummed
strong-field limit based on scale identification. As our flow equation approach
does not need scale-identification arguments and goes beyond the leading-log
resummation, we consider the existence of a global Lagrangian as being in line
with the fact that magnetic fields do not transfer energy to charged
fluctuations; therefore the strong magnetic field limit can behave differently
from the high-momentum limit of the theory. 

We emphasize that our observation of fixed points in pure nonlinear
electrodynamics does not resolve the triviality problem of QED. The latter is
tightly linked to charged particle fluctuations not being part of our pure
abelian gauge theory setting. Also, we do not observe an immediate mechanism
that could balance the charged fluctuations within nonlinear electrodynamics.
Still, the family of non-Gaussian fixed points observed in this work could play
a useful role in models with gauge-kinetic or nonminimal interactions to other
particle sectors, potentially contributing to mechanisms of UV completion.

%
%
%
\section*{ACKNOWLEDGMENTS}

We thank Gustavo de Brito, Astrid Eichhorn, Benjamin Knorr, Roberto Percacci, Abdol Sabor Salek,
Marc Schiffer, Inti Sodemann Villadiego, and Luca Zambelli for valuable discussions.  
This work has been funded by the Deutsche 
Forschungsgemeinschaft (DFG) under Grant No. 406116891 
within the Research Training Group RTG 2522/1 and under Grants No.
392856280, No. 416607684, and No. 416611371 within the  Research Unit FOR2783/2.

\bibliography{bibliography}

\end{document}